\documentclass{IEEEojcsys}

\usepackage{cite}
\usepackage{amsmath,amssymb,amsfonts}
\usepackage{graphicx}
\usepackage{textcomp}
\usepackage{algorithm2e}
\usepackage{dirtytalk}
\makeatletter
\renewcommand{\@algocf@capt@plain}{above}
\newcommand{\specialcell}[1]{\ifmeasuring@#1\else\omit$\displaystyle#1$\ignorespaces\fi}
\makeatother
\usepackage{cuted}
\usepackage{color,array}
\usepackage{mathtools}
\mathtoolsset{showonlyrefs}
\usepackage{amsthm}
\let\labelindent\relax
\usepackage{nohyperref}
\usepackage{enumitem}

\newtheorem{theorem}{Theorem}[section]
\newtheorem{definition}{Definition}[section]
\newtheorem{assumption}{Assumption}[section]
\newtheorem{problem}{Problem}

\newtheorem{corollary}{Corollary}[theorem]
\newtheorem{lemma}[theorem]{Lemma}
\newtheorem{prop}{Proposition}[section]
\newtheorem{remark}{Remark}
\newtheorem{exmp}{Example}[section]
\usepackage{graphicx}

\jvol{00}
\jnum{XX}
\paper{1234567}
\pubyear{2023}
\receiveddate{XX XX 2023}
\accepteddate{XX XX 2023}
\publisheddate{XX XX 2023}
\currentdate{XX XX 2023}
\doiinfo{OJCSYS.2023.Doi Number}

\renewcommand\thesubsection{\thesection.\Alph{subsection}}
\begin{document}

\sptitle{Article Category}

\title{Risk-based Security Measure Allocation Against Actuator attacks
} 
\editor{This paper was recommended by Associate Editor F. A. Author.}
\author{Sribalaji C. Anand \affilmark{1} (Student Member, IEEE)}

\author{Andr\'e M. H. Teixeira \affilmark{2}  (Member, IEEE)}


\affil{Department of Electrical Engineering, Uppsala University, PO Box 65, SE-75103, Uppsala, Sweden.} 
\affil{Department of Information Technology, Uppsala University, PO Box 337, SE-75105, Uppsala, Sweden.} 

\corresp{CORRESPONDING AUTHOR: Sribalaji C. Anand (e-mail: {sribalaji.anand@angstrom.uu.se})}
\authornote{This work is supported by the Swedish Research Council grant 2018-04396 and by the Swedish Foundation for Strategic Research.}

\markboth{Risk-based Security Measure Allocation Against Injection Attacks on Actuators}{Sribalaji C. Anand {\itshape et. al}.}

\begin{abstract}
This article considers the problem of risk-optimal allocation of security measures when the actuators of an uncertain control system are under attack. We consider an adversary injecting false data into the actuator channels. The attack impact is characterized by the maximum performance loss caused by a stealthy adversary with bounded energy. Since the impact is a random variable, due to system uncertainty, we use Conditional Value-at-Risk (CVaR) to characterize the risk associated with the attack. We then consider the problem of allocating security measures to the set of actuators to minimize the risk. We assume that there are only a limited number of security measures available. Under this constraint, we observe that the allocation problem is a mixed-integer optimization problem. Thus we use relaxation techniques to approximate the security allocation problem into a Semi-Definite Program (SDP). We also compare our allocation method $(i)$ across different risk measures: the worst-case measure, the average (nominal) measure, and $(ii)$ across different search algorithms: the exhaustive and the greedy search algorithms. We depict the efficacy of our approach through numerical examples.
\end{abstract}

\begin{IEEEkeywords}
Networked control systems, Resilient Control Systems, LMIs, Optimization.
\end{IEEEkeywords}

\maketitle

\section{INTRODUCTION}\label{sec:introduction}
Security of Networked Control Systems (NCSs) has received increased research attention \cite{sandberg2022secure,dibaji2019systems}. Following \cite{chong2019tutorial}, the literature on the security of NCSs can be broadly classified into (i) characterizing the different attack scenarios, (ii) determining the optimal attack strategies and their corresponding impact (performance loss), and (iii) attack mitigation.

In the literature, attack mitigation (defined in \cite[Chapter 1]{ferrari2021safety}) is performed (mostly) in three methods. The first method is to design mechanisms to detect attacks \cite{giraldo2018survey,li2023attack}. The second method is to design the parameters of the closed-loop system (controller gain, for instance) so that the attack impact is minimal \cite{hashemi2019co,hespanha2019output,fotiadis2022concurrent}. The third method is to allocate the security measures (encryption for instance) so that the attack impact through the unprotected assets is minimal \cite{milovsevic2020security}. 

\subsection{Literature review}

The problem of security allocation in NCS can be interpreted in different ways. Here we provide three different types of security allocation which can also be found in the literature. Firstly at the device level, consider patching of Programmable Logic Controllers (PLC) in NCSs. In contrast to classical Information Technology (IT) systems, NCSs have strict real-time requirements which require the operator to carefully consider the risks involved before deploying the security patch \cite{tom2008recommended}. The risk can vary from exposing additional vulnerabilities \cite{beattie2002timing}, increasing the downtime, operational costs, etc. Then allocation refers to the problem of deciding which PLCs to be updated with a patch by not enduring a huge performance loss in the presence of attacks. 

Secondly, similar to classical IT systems, security allocation can refer to deciding which of the communication channels to encrypt (at the network communication level) \cite{risley2003electronic}. Here, encryption can refer to message authentication or private key encryption which can be implemented in NCS without any significant delays (see \cite{gomez2018operator} and \cite{hadley2007secure}).

Finally, security allocation at the control level refers to choosing the communication channels to add watermarks. There are many strategies in the literature, designed from a control perspective, so that the adversary cannot remain stealthy \cite{mo2009secure,ferrari2020switching,du2021secure}. It is to be noted that the allocation of such control-theoretic strategies for attack detection does not introduce significant delays in the closed-loop system.

The problem of security allocation has been studied extensively in the literature. For instance, the security allocation strategy against sensor attacks in static power systems was investigated in \cite{dan2010stealth}. The work \cite{milovsevic2020security} focuses on security allocation for deterministic dynamical systems by exploiting sub-modularity. The optimal sensor/detector placement problem for uncertain systems was studied in \cite{nguyen2022zero} using a game-theoretic approach. However, there are three main differences between \cite{nguyen2022zero} and our work. Firstly, \cite{nguyen2022zero} considers a framework with a single attacked node whereas we consider multiple attacked actuators. Secondly, \cite{nguyen2022zero} requires explicitly calculating the game payoff for each pair of players' actions, whereas we propose a single-shot semi-definite program (SDP). Finally, \cite{nguyen2022zero} uses the Value-at-Risk (VaR) as the risk metric, whereas we use Conditional VaR (CVaR) as a risk metric, which has much more advantages, notably convexity \cite{rockafellar2000optimization}. We note that game-theoretic approaches have also been used in other research fields not limited to control for optimal allocation of monitoring resources  \cite{krause2011randomized,rahmattalabi2018robust}

Other works which focus on optimal security allocation using a game-theoretic approach are \cite{pirani2021game,pirani2021strategic,milosevic2023strategic,milovsevic2019network}. However, there are three main differences to our approach. Firstly, these works focus on sensor attacks, whereas we focus on actuator attacks. Secondly, these works do not consider uncertain systems but they focus on large-scale systems and their related complexity issues. Finally, these works focus on attack detection only but do not consider the related performance loss caused by attacks. In contrast, we consider a metric that takes into account the performance loss as well as the detection performance; see \cite{anand2020joint} and \cite[Chapter 4]{ferrari2021safety}. One of our previous works \cite{anand2022risk_acc} also uses a similar metric for optimizing CVaR, however, it focuses on controller design rather than security allocation. Finally, another of our previous works \cite{anand2022risk} focuses on allocation for uncertain systems. However, \cite{anand2022risk} is based on an exhaustive search.

\subsection{Problem setup}
\begin{figure}
    \centering
    \includegraphics[width=18.5pc]{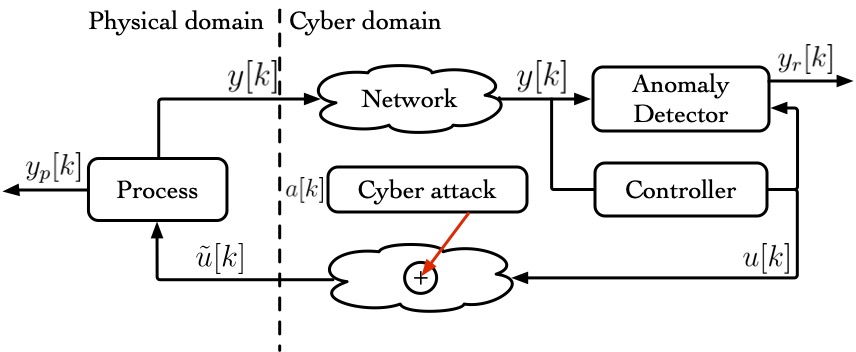}
    \caption{NCS under false data injection attack on actuators}
    \label{System}
\end{figure}

As mentioned in the literature study, the allocation problem for uncertain systems has not been studied yet. Nonetheless, uncertainties are inevitable in physics-based \cite{tan2009unified} or data-based modeling techniques \cite{verhaegen2007filtering}. To address this gap, we consider an uncertain linear time-invariant process \eqref{P}. Since the process is controlled with a feedback controller \eqref{C} over a wireless network, it is prone to cyber-attacks. Thus, we consider false data injection attacks on the actuators and an observer-based detector \eqref{D}. The closed loop system under attack is described in \eqref{P}-\eqref{D} (also see Figure \ref{System})
\begin{align}
    \mathcal{P}: & \left\{
                \begin{array}{ll}
                    \dot{\bar{x}}(t) &= A^{\Delta}\bar{x}(t) + B^{\Delta}{\tilde{u}(t)} \\
                    y(t) &= C\bar{x}(t)\\
                    y_p(t) &= C_j\bar{x}(t)
                 \end{array}
                \right. \label{P}\\
    \mathcal{C}: & \left\{
                 \begin{array}{ll}
                     \dot{z}(t) &= A_cz(t)+ B_cy(t)\\
                     u(t) &= C_cz(t)+ D_cy(t)
                \end{array}
                \right. \label{C}
                                \end{align}
                \begin{align}
    \mathcal{D}: & \left\{
                \begin{array}{ll}
                \dot{\hat{x}}_p(t) &= A\hat{x}_p(t) + B{u}(t) + K y_r(t)\\
                    y_r(t) &= y(t)-C\hat{x}(t)\\
                \end{array}
                \right. \label{D}
\end{align}
where  $A^{\Delta} \triangleq A + \Delta A(\delta)$ with $A$ representing the nominal system matrix, and the parametric uncertainty characterized by $\Delta A(\delta), \delta \in \Omega$. We assume $\Omega$ to be closed, bounded, and to include the zero uncertainty yielding $\Delta A(0) = 0$. The other matrices are similarly expressed. The state of the process, controller, and detector is represented by $\bar{x}(t) \in \mathbb{R}^{n_x}, z(t) \in \mathbb{R}^{n_z}$ and $ \hat{x}_p(t) \in \mathbb{R}^{n_x}$ respectively. The control signal generated by the controller and the control signal received by the process is $ u(t) \in \mathbb{R}^{n_u}$ and $\tilde{u}(t) \in \mathbb{R}^{n_u}$ respectively. The measurement output, performance output, and residue output are denoted by $y(t) \in \mathbb{R}^{n_m}, y_p(t) \in \mathbb{R}^{n_p}$ and $y_r(t) \in \mathbb{R}^{n_m}$ respectively. 

We consider an adversary with complete system knowledge injecting false data into the actuator channel. The operator is the only one with uncertainties in system knowledge. This setup might be unrealistic, but it helps us study the worst-case scenario. The main problem investigated in this article is formulated next. 
\begin{problem}\label{problem_00}
Given the uncertain NCS is under attack, and that we can secure a small number of actuators ($n_w$), how to optimally allocate the security measures?$\hfill\triangleleft$
\end{problem}
\subsection{Contributions}
To the best of the authors' knowledge, the problem of security allocation for uncertain systems using CVaR has not been addressed in the literature. To this end, the main contributions of the article are as follows
\begin{enumerate}
\item We formulate the optimal allocation problem for uncertain systems using CVaR as a risk metric. The attack impact is characterized by the maximum performance loss caused by a stealthy adversary with limited energy. 
\item The impact metric is non-convex in the design variable. Hence we derive a convex \textit{impact proxy} which also serves as the upper bound for the impact.
\item The allocation problem using CVaR and the proposed impact proxy is a mixed integer SDP which is generally hard to solve \cite{gally2018framework}. Through relaxations, we propose an approximate SDP to solve the allocation problem, along with posterior guarantees on the optimality gap.
\item We compare our solution across different risk measures (worst-case, and nominal measures) and different search algorithms (exhaustive, and greedy search).
\end{enumerate} 

The rest of this article is organized as follows: We formulate the problem in Section \ref{sec:PF}. We propose a convex SDP to solve Problem \ref{problem_00} approximately in Section \ref{sec:SDP}. We outline the solution to the security allocation problem under different risk measures in Section \ref{sec:worst}. We describe the exhaustive search algorithm and greedy search algorithm in Section \ref{sec:exhaust:greedy}, where we also compare the methods briefly. We depict the efficacy of our proposed approach through numerical examples in Section \ref{sec:NE} and conclude in Section \ref{sec:Conclusion}. 

\subsection{Notation}
A positive (semi-) definite matrix $A$ is denoted by $A \succ 0\;(A \succeq 0)$. Let $x: \mathbb{R} \to \mathbb{R}^n$ be a continuous-time signal with $x(t)$ as the value of the signal $x$ at the time $t$. Let the time horizon be $[0,N]=\{ t \in \mathbb{R}^+|\; 0 \leq t\leq N \}$. The $\mathcal{L}_2$-norm of $x$ over the horizon $(0,N)$ is represented as $|| x ||_{\mathcal{L}_2, (0,N)}^2 \triangleq \int_{t=0}^{N} x(t)^Tx(t)$. Let the space of square integrable signals be defined as $\mathcal{L}_2 \triangleq \{ x: \mathbb{R}^+ \to \mathbb{R}^n |\; ||x||^2_{\mathcal{L}_2, (0,\infty)} < \infty\}$ and the extended signal space be defined as $\mathcal{L}_{2e} \triangleq \{ x: \mathbb{R}^+ \to \mathbb{R}^n | \;||x||^2_{(0,N)} < \infty, \forall N \in \mathbb{R}^+ \}$. For the sake of simplicity, we represent $||x||^2_{\ell_2,(0,\infty)}$ as $||x||^2_{\ell_2}$. Given a vector $f \in\mathbb{R}^N$, let $\{x_i\}$ be the $N$ indices of entries of $f$ such that $f(x_1) \geq f(x_2) \geq \dots \geq f(x_N)$. Then, the $(N-i+1)$-th order statistic of $f$ is given by $f(x_i)$.  And $f^{[j]}, j\leq N$ represents the $j$-th element of the vector $f$.
\section{Problem Formulation}\label{sec:PF}
The system \eqref{P}-\eqref{D} is said to have a good performance when $||y_p||_{\ell_2}^2$ is small. This is similar to linear quadratic (LQ) control where the objective is to minimize performance loss. Similarly, an anomaly is considered to be detected when the detector output energy $||y_r||_{\ell_2}^2$ is greater than a predefined threshold, say $\epsilon_r$. Given this setup, we next describe the adversary in detail and later formulate the problem.
\subsection{Disruption and disclosure resources}\label{disclosure:sec} 
The adversary can access (eavesdrop) the control channels and can inject data. This is represented by $$\tilde{u}(t) \triangleq u(t) + B_a a(t)$$ where $a(t) \in \mathbb{R}^{n_u}$ is the data injected by the adversary. The matrix $B_a$ is a diagonal matrix with $B_a(i,i)=1$ if the actuator channel $i$ is under attack and zero otherwise. The matrix $B_a$ is square, however, this does not enforce the adversary to attack all the actuators. If the adversary is interested in attacking some of the actuators, the adversary can simply set the corresponding attack vector to zero. 

In general, $B_a$ is chosen by the operator for analysis purposes. If the operator believes that the actuator channel (say $j$) might be under attack, then the corresponding channel has an entry 1 ($B_a(j,j)=1$). In the rest of the article, the matrix $B_a$ is called the \textit{attack matrix}.
\subsection{Attack goals and constraints.}
The adversary's objectives are contrary to that of the operator. That is, the adversary aims to disrupt the system's behavior while staying stealthy. The system disruption is evaluated by the increase in energy of the performance output, whereas the adversary is stealthy if the energy of the detection output is below a predefined threshold (namely $\epsilon_r$). 

In reality, the adversary stops attacking the system after some unknown time $T<\infty$. Additionally, the corrupted input signal is applied by physical actuators which have actuator bounds. Thus we consider the energy of the attack signal to be bounded by a predefined threshold (namely $\epsilon_a$).
\subsection{System knowledge} 
Next, we consider that the adversary has full system knowledge, i.e., s/he knows the system matrices \eqref{P}-\eqref{D}. We define such an adversary as an omniscient adversary.
\begin{definition}[Omniscient adversary]\label{OA}
An adversary is defined to be omniscient if it knows the matrices in \eqref{system:CL:uncertain}.$\hfill\triangleleft$
\end{definition}
 
In reality, it is hard to know the system matrices of \eqref{system:CL:uncertain} due to uncertainty. Thus, such an adversarial setup is far from reality but can help study the worst case. Readers interested in realistic setups where the adversary also has uncertainty are referred to \cite{anand2021risk}. However, as mentioned in \cite{anand2021risk}, analysis of such realistic setups is computationally intensive. Thus, in this article, we focus on the omniscient adversary. 

Defining ${x}(t) \triangleq [ x_p(t)^T \; z(t)^T\;\hat{x}_p(t)^T]^T$, the closed-loop system under attack with the performance output and detection output as system outputs becomes
\begin{equation}\label{system:CL:uncertain}
                \begin{array}{ll}
                            \dot{x}(t) &= {A}_{cl}^{\Delta}{x}(t) + {B}_{cl}^{\Delta}a(t),\\
                            y_p(t) &= {C}_p{x}(t),\\ y_r(t) &= {C}_r{x}(t),\\
                \end{array}
\end{equation}
with 
$\left[ \begin{array}{c|c}
A_{cl}^{\Delta} & B_{cl}^{\Delta}
\end{array}\right] = $ 
\begin{align*}
&\left[
\begin{array}{c|c}
\begin{matrix}
    A^{\Delta}+B^{\Delta}D_cC & B^{\Delta}C_c & 0\\
    B_cC & A_c & 0\\
    (BD_c +K_e)C & BC_c & A-KC
    \end{matrix} & \begin{matrix}
    B^{\Delta}B_a \\0 \\ 0 
    \end{matrix}
\end{array}\right]\\
&{C}_p \triangleq \begin{bmatrix}
    C_j & 0 & 0
    \end{bmatrix},\;\text{and}\; {C}_r \triangleq \begin{bmatrix}
    C & 0 & -C
    \end{bmatrix}.
\end{align*}

In \eqref{system:CL:uncertain}, the signals $x, y_p,$ and $y_r$ are also functions of uncertainty, and the superscripts are dropped for simplicity. Next, we establish the following assumptions.
\begin{assumption}\label{assume_stable}
Closed-loop system \eqref{system:CL:uncertain} is stable $\forall \delta \in \Omega$.$\hfill\triangleleft$
\end{assumption}
\begin{assumption}\label{Ass:minimal}
The tuple $(A_{cl}^{\Delta},B_{cl}^{\Delta})$ is controllable $\forall \delta \in \Omega$. The tuple $(A_{cl}^{\Delta},\begin{bmatrix}
    C_p^T & C_r^T
\end{bmatrix}^T)$ is observable $\forall  \delta \in \Omega$. $\hfill\triangleleft$
\end{assumption}
Assumption \ref{assume_stable} states that the feedback controller robustly stabilizes the plant. Assumption \ref{Ass:minimal} is a direct consequence (and a common assumption in dissipative systems theory) of \cite{trentelman1991dissipation} which is later used to formulate the proof of Lemma \ref{thm0}. We later also briefly discuss the consequence of relaxing Assumption \ref{Ass:minimal} (See Remark \ref{rem:minimal}).
\subsection{Optimal allocation problem}
Consider the data injection attack scenario where the parametric uncertainty $\delta \in \Omega$ of the system is known to the adversary but not to the defender. Under this setup, the adversary can cause high disruption by remaining stealthy as it will be able to inject attacks by solving \eqref{eqq4},
\begin{equation}\label{eqq4}
\begin{aligned}
q(B_a,\delta)  \triangleq \sup_{a\in \mathcal{L}_{2e}} \quad & \Vert y_p[B_a,\delta] \Vert_{\mathcal{L}_2}^2 \\
\textrm{s.t.} \quad & \Vert y_r[B_a,\delta] \Vert_{\mathcal{L}_2}^2 \leq \epsilon_r\\
& \Vert a[\delta] \Vert_{\mathcal{L}_2}^2 \leq \epsilon_a,\;x[B_a,\delta](0)=0,
\end{aligned}
\end{equation}
where $y_p[{B_a,\delta}], y_r[{B_a,\delta}],$ and $a[\delta]$ are the performance output, detection output, and the attack vector corresponding to the matrix $B_a$ and uncertainty $\delta$, and $q(\cdot)$ is the impact caused by the adversary on \eqref{system:CL:uncertain}. Such a setup in \eqref{eqq4} is considered for the adversary, to analyze the worst-case impact of stealthy attacks, since the adversary will be able to inject undetectable attacks which cause high performance deterioration. For the defender, $q(B_a,\delta)$ becomes a random variable since $\delta$ is unknown. The defender only knows the bounds of the set $\Omega$, the nominal system matrices in \eqref{system:CL:uncertain}.

Thus, the defender protects some of the actuators (through encryption for example) such that the risk corresponding to $q(\cdot)$ in \eqref{eqq4} is minimized. However, the defender also has the constraint that there are only a limited number of security measures i.e., $n_w < n_u$ ($C1$). Recall that the diagonal entries of the matrix $B_a$ can either be $1$ (unprotected) or $0$ (protected) ($C2$). Then \textit{Problem \ref{problem_00}} can be re-formulated as
\begin{problem}\label{problem_0}
Find the optimal diagonal matrix $B_a^*$ such that
\begin{equation}\label{eq:problem_0}
\begin{aligned}
B_a^* \triangleq \arg \inf_{B_a}\quad & \mathcal{R}_{\Omega}(q(B_a,\delta)) & \\
\text{s.t.}\quad &  \sum_{i=1}^{n_u} B_a(i,i) \geq n_u-n_w, &  (C1) \\
& B_a(i,i) = \{0,1\} & (C2)
\end{aligned}
\end{equation}
where $\mathcal{R}_{\Omega}$ is a risk metric chosen by the defender. The subscript $\Omega$ denotes that the risk acts over the set $\Omega$ whose probabilistic description is known to the defender (for the results of this article to hold, it is sufficient that the defender can draw samples from the set $\Omega$). 
$\hfill\triangleleft$
\end{problem}

CVaR is extensively used in the literature due to its numerous advantages \cite{rockafellar2000optimization}. Thus we choose the CVaR as a risk metric in \textit{Problem \ref{problem_0}}. {Before we introduce the risk metric, we make the following assumptions that follow from \cite{rockafellar2000optimization}.}
\begin{assumption}
The defender can draw samples from the set $\Omega$ and the function $q(\cdot,\delta)$ is continuous. $\hfill\triangleleft$
\end{assumption}
\begin{definition}\label{def_Var}
Given a random variable $q(\cdot,\delta)$ with density $p(q)$, the $\text{CVaR}_{\alpha}(q(\cdot,\delta))$ (given $\alpha \in (0,1)$) is given by
\begin{equation}\label{cvar_equation}
\frac{1}{1-\alpha}\int_{q(\cdot,\delta)|q(\cdot,\delta) \geq \text{VaR}_{\alpha}\{q(\cdot,\delta)\}} q(\cdot,\delta)\; p(q)\; dq
\end{equation}
where
$\text{VaR}_{\alpha}\{q(\cdot,\delta)\} \triangleq \inf \{x|\mathbb{P}_{\Omega}[q(\cdot,\delta) \leq x] \geq 1-\alpha\}$
$\hfill\triangleleft$
\end{definition}

{Next, we illustrate the risk metrics through an example, whereby also motivating the choice of the risk metric.}

\begin{exmp}\label{ex_first}
Consider the system in \eqref{P}-\eqref{D} where
\begin{align}\label{matrix_NE}
\left[
\begin{array}{c|c}
A^{\Delta} & C^T
\end{array}
\right] &=
\left[ 
\begin{array}{c|c}
\begin{matrix}
-1 &0& 0 &\delta\\1 &-5 &0& 0\\1 &1 &-9 &0\\ 10& 1& 10& -1
\end{matrix}     & \begin{matrix}
1 \\ 0 \\ 1 \\0
\end{matrix}
\end{array}
\right]\\
\left[
\begin{array}{c}
-L \\ \hline K \\ \hline C_j
\end{array}
\right] &= 
\left[ 
\begin{array}{c}
\begin{matrix}
5.26 & 0.44 & 1.64 & 1.99\\
0.44 & 0.13 & 0.14 & 0.17\\
1.64 & 0.14 & 0.61 & 0.68\\
1.99 & 0.17 & 0.68 & 0.87
\end{matrix} \\
\hline \begin{matrix}
5.70 & 0.70 & 0.55 & 15.28
\end{matrix}\\
\hline \begin{matrix}
\;\;1\;\; & \;\;1\;\; & \;\;1\;\; & \;\;1\;\;
\end{matrix}
\end{array}
\right],
\end{align} 
$\delta \in \Omega \triangleq [0,\;3],$and $ B=C_j^T$. We set $\epsilon_r=1,$ and $\epsilon_a=300$. We determine the value of the random variable $q(\cdot,\delta)$ in \eqref{eqq4} for different uncertainty realizations $\delta \in \Omega$ and plot the probability density function of $q(\cdot)$ in Figure \ref{fig:var_cvar_wc}. We depict the value of the risk measures: VaR$_{0.1}\{q(\cdot)\}$, CVaR$_{0.1}\{q(\cdot)\}$, $\mathbb{E}\{q(\cdot)\}$, worst-case (similar to $H_{\infty}$ control \cite{petersen2012robust}), and nominal measure (without considering uncertainties). Detailed definitions of worst-case and nominal measures are given in Section \ref{sec:WC} and \ref{sec:NC} respectively.

\begin{figure*}
    \centering
    \includegraphics[scale=0.3]{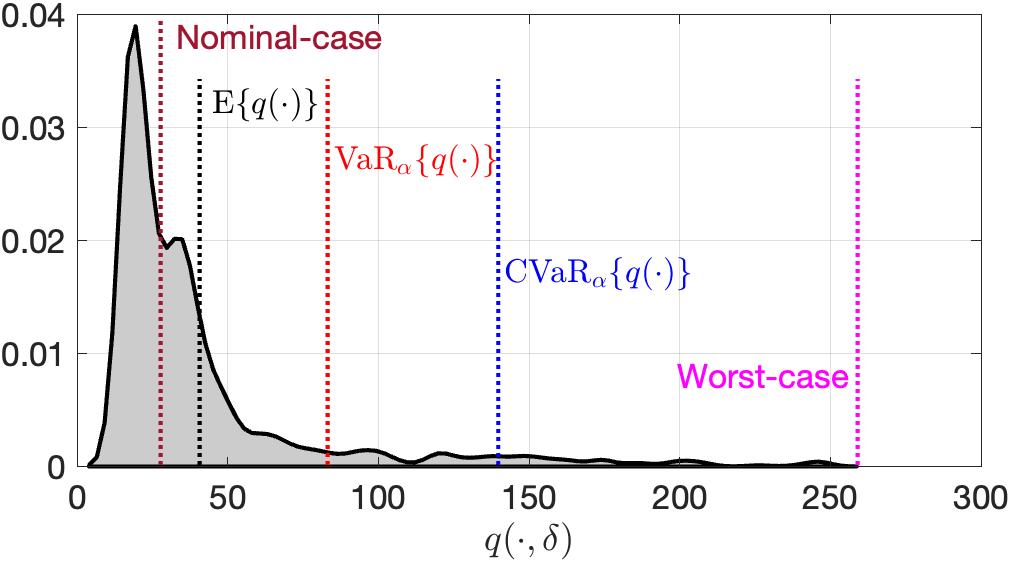}
    \caption{Probability distribution function of the random variable $q(\cdot,\delta)$ and the corresponding risk measures.}
    \label{fig:var_cvar_wc}
\end{figure*}

Let $x=\text{VaR}_{\alpha}\{q(\cdot)\}$. Then by optimizing the VaR, one optimizes the probability that the value of $q(\cdot) \geq x$. However, VaR does not take into account the thickness of the tail pdf of $q(\cdot)$. In general, although we want the risk of attacks to be minimal, we allow for events whose probability is very low but with a {high} impact. In such scenarios, optimizing the worse case measure might be conservative. The nominal measure is also conservative since it does not consider uncertainties. Given the above arguments, we choose CVaR as the risk metric in this article. $\hfill \triangleleft$
\end{exmp}

In our setting, the defender determines the attack matrix $B_a$ such that $\text{CVaR}_{\alpha}q(B_a,\delta)$ (given $\alpha$) is minimized. To this end, \textit{Problem 2} can be reformulated as
\begin{equation}\label{problem_1}
B_a^* = \arg\inf_{B_a} \left\{\text{CVaR}_{\alpha}\{q(B_a,\delta)\}\Big| (C1),(C2)\right\}.
\end{equation}

Although CVaR is a convex function, there are three difficulties in solving \eqref{problem_1}. Firstly, $q(\cdot)$ is non-convex in the design variable $B_a$, which we address in Section \ref{sec:SDP_1}. Secondly, the design variable $B_a$ is binary $(C2)$ which makes the design problem non-convex, and we address this issue in Section \ref{sec:SDP_3}. Finally, CVaR cannot be efficiently evaluated exactly since $\Omega$ is continuous. We describe an empirical approach to solve \eqref{problem_1} in Section \ref{sec:SDP_2}. Before we discuss the solution to \eqref{problem_1}, we briefly discuss the relation between \eqref{eqq4} and other attack impact metrics in the literature.
\begin{remark}[Relation between Problem \ref{problem_0} and a {Stackelberg} game]
Problem 2 can be related to a {Stackelberg} game \cite{pirani2021game} as follows. The defender first selects the action ($B_a$), i.e., which actuators to protect. Then, the adversary selects the attack $(a\in \ell_{2e})$ through the unprotected actuators and computes the optimal attack by solving (5). Thus Problem 2 can also be viewed from a game theoretic point of view. $\hfill\triangleleft$
\end{remark}
\begin{remark}[Boundedness of risk metric]\label{rem_1}
The concept of risk is sensible when it is finite. In our setup, the risk is finite if the random variable $q(\cdot)$ is finite. Thus, in the rest of the sequel, we assume that $q(\cdot,\delta)$ is bounded $\forall \delta \in \Omega$. Since the closed loop system is stable, the value of $q(\cdot,\delta)$ in \eqref{eqq4} is unbounded iff $||y_p[\cdot,\delta]||_{\mathcal{L}_2}^2$  is unbounded, which consequently is unbounded iff $||a[\delta]||_{\mathcal{L}_2}^2$ is unbounded. But we know that $||a[\delta]||_{\mathcal{L}_2}^2 \leq \epsilon_a$ where $\epsilon_a$ is bounded. Thus the assumption on the boundedness of the random variable $q(\cdot)$ is logical.$\hfill\triangleleft$
\end{remark}
\begin{remark}\label{rem_2}
When $\alpha \to 1$, the optimization problem \eqref{problem_1} minimizes the worst case impact \eqref{eqq4} across all uncertainties. However, as explained in Example \ref{ex_first} and \cite{muller2018risk}, such approaches can be conservative.
$\hfill\triangleleft$
\end{remark}
\subsection{Relation between (\ref{eqq4}) and other metrics}\label{sec:discuss}
In this article, for any given uncertainty $\delta \in \Omega$, we use \eqref{eqq4} to capture the amount of disruption caused by the adversary. However, other security metrics in the literature can be related to the metric \eqref{eqq4}.

Let $\epsilon_r \gg \epsilon_a$. That is, the detection threshold becomes very large that the constraint on the detection output becomes inactive. Then \eqref{eqq4} becomes the $H_{\infty}$ metric where the attack is treated as the disturbance. An SDP to determine the ${H}_{\infty}$ metric can be found in \cite[(6)]{hilhorst2014extended}. {Works such as \cite{bopardikar2016h,you2015convex}} for instance, use the $H_{\infty}$ metric for measuring attack impact.  

{On the other hand}, when $\epsilon_a \gg \epsilon_r$, the constraint on the attack energy becomes inactive. Then \eqref{eqq4} is the Output-to-Output Gain (OOG) \cite[Chapter 6]{ferrari2021safety}. OOG has many advantages over the $H_{\infty}$ and $H_{\_}$ metric which we discussed in \cite{anand2020joint}. An SDP to determine the OOG can be found in \cite[(6.18)]{ferrari2021safety}. We combine the above results in Proposition \ref{prop2}. 
\begin{prop}\label{prop2}
Consider the CT system under attack described in \eqref{system:CL:uncertain} and the corresponding impact metric described in \eqref{eqq4}. Then, given $\delta \in \Omega$, the following statements are true.
\begin{enumerate}
    \item Let $\epsilon_a$ be a constant, and let $\gamma_a$ represent the classical $H_{\infty}$ gain of the closed loop system \eqref{system:CL:uncertain} for a given $\delta \in \Omega$. Then it holds that $\lim_{\epsilon_r \to \infty} q(B_a,\delta) = \gamma_a\epsilon_a$.
    \item Let $\epsilon_r$ be a constant, and let $\gamma_r$ represent the OOG \cite[(6.18)]{ferrari2021safety} of closed-loop system \eqref{system:CL:uncertain} for a given $\delta \in \Omega$. Then it holds that $\lim_{\epsilon_a \to \infty} q(B_a,\delta) = \gamma_r\epsilon_r$. $\hfill\square$
\end{enumerate}
\end{prop}

The objective of the exercise in Proposition \ref{prop2} is to show that the allocation results in this article, which are based on the metric \eqref{eqq4}, can be related to other results (based on $H_{\infty}$ metric or OOG) by varying the value of $\epsilon_r$ and $\epsilon_a$. In the next section, we start to solve \eqref{problem_1}.
\section{Convex SDP for optimal allocation}\label{sec:SDP}
In this section, we first consider a sampled uncertainty $\delta_i$ and show that, given $B_a$, the value of $q(B_a,\delta_i)$ can be determined via a convex SDP. We also show that the SDP is a non-convex function of the design variable $B_a$. Then we propose a relaxed SDP which is convex in $B_a$. We later use this relaxed SDP, to formulate a convex allocation problem.
\subsection{Convex relaxation for the impact metric}\label{sec:SDP_1}
Let us consider the impact metric $q(B_a,\delta_i)$ in \eqref{eqq4}. We show in Lemma \ref{lem0} that its value can be determined by its convex dual (the proof of Lemma \ref{lem0} and all the other results in the sequel are presented in the Appendix).
\begin{lemma}\label{lem0}
Given a sampled uncertainty $\delta_i$, and an attack matrix $B_a$, the value of the impact $q(B_a,\delta_i)$ can be calculated by its convex dual counterpart \eqref{dual} where $\gamma_1$ and $\gamma_2$ are the Lagrange multipliers of the constraints.
\begin{equation}\label{dual}
    \begin{aligned}
    \inf_{\gamma_{1,i},\gamma_{2,i}} \quad & \epsilon_r\gamma_{1,i} + \epsilon_a\gamma_{2,i} \\
\text{s.t.} \;\;\quad & \Vert y_p[B_a,\delta_i] \Vert_{\mathcal{L}_2}^2 - \gamma_1 \Vert y_r[B_a,\delta_i]\Vert_{\mathcal{L}_2}^2\\
& \qquad \qquad \qquad -\gamma_2 \Vert a[\delta_i] \Vert_{\mathcal{L}_2}^2 \leq 0, \forall a \in \mathcal{L}_{2e} \\
& x[B_a,\delta_i](0)=0, \gamma_{1,i} \geq 0,\gamma_{2,i} \geq 0. 
\end{aligned}
\end{equation}
\end{lemma}

Although \eqref{dual} is convex, it is hard to solve \eqref{dual} since the constraints lie in the signal space. Thus, we use dissipative system theory to re-write \eqref{dual} as a convex SDP. 

Before we formulate this SDP, we introduce the following notation. The matrices in \eqref{system:CL:uncertain} under a sampled uncertainty $\delta_i$ is denoted as $A_{cl,i},B_{cl,i},C_{p}$ and $C_{r}$. Correspondingly the signals under the sampled uncertainty $\delta_i$ becomes $a_i,y_{p,i},y_{r_i}$ and $x_i$. We also know from \eqref{system:CL:uncertain} that $B_{cl,i}^T$ takes the form $\begin{bmatrix}
    B_a^TB_i^T & 0 & 0
\end{bmatrix}$ and thus is a linear function in $B_{a}$.
\begin{lemma}\label{thm0}
Let Assumption \ref{Ass:minimal} hold. Then, for a sampled uncertainty $\delta_i$, the optimization problems \eqref{dual} and \eqref{o2} are equivalent.
\begin{equation}\label{o2}
\begin{aligned}
\min_{\mathcal{S}_{1}} \quad & \epsilon_r\gamma_{1,i} + \epsilon_a\gamma_{2,i} \\
\textrm{s.t.}\quad & \begin{bmatrix}
\mathcal{W}_i&P_iB_{cl,i}\\B_{cl,i}^TP_i&-\gamma_{2,i}I
\end{bmatrix}\preceq 0 & (C3_i)
\end{aligned}
\end{equation}
where $\mathcal{S}_1 \triangleq \{P_i \succ 0,\gamma_{1,i} \geq 0,\gamma_{2,i} \geq 0\},$ and $\mathcal{W}_i=A_{cl,i}^TP_i+ P_iA_{cl,i} +C_{p}^TC_{p} -\gamma_{1,i} C_{r}^TC_{r}. \hfill \square$
\end{lemma}

Lemma \ref{thm0} proposes an SDP to determine $q(\cdot)$ under a sampled uncertainty. However, \eqref{o2} is non-convex in $B_a$ as $(C3_i)$ contains the term $P_{i}B_{cl,i}$ which is bi-linear (since $B_{cl,i}$ is a linear function of $B_a$). Thus, we propose a relaxed SDP in place of \eqref{o2} which is convex in $B_a$. Henceforth, the value of this relaxed SDP is denoted by $\tilde{q}(\cdot)$. 

The main objective of proposing this relaxed SDP is: once we show that $\tilde{q}(\cdot)$ is a convex function of the design variable $B_a$, we can substitute this convex function $ \tilde{q}(\cdot)$ into the definition of CVaR in \eqref{problem_1} (replacing the non-convex function $q(\cdot)$) and optimize it. Now we state our main result. 
\begin{theorem}\label{thm1}
Given a sampled uncertainty $\delta_i$, the SDP
\begin{equation}\label{sdp1}
\begin{aligned}
\min_{\mathcal{S}_2}\;&\epsilon_r\gamma_{1,i} + \epsilon_a\gamma_{2,i} \\
\textrm{s.t.}\;&\begin{bmatrix}
-I & 0 & C_{p}X_i & 0\\ 0 & -\gamma_{1,i}I & I & 0\\ X_iC_{p}^T & I & \mathcal{W}_{2,i} & B_{cl,i}\\0 & 0 & B_{cl,i}^T & -\gamma_{2,i}I \end{bmatrix} \preceq 0 & (C4)\\
&\mathcal{S}_2 = \{X_i \succ 0,\;\gamma_{1,i} \geq 0,\;\gamma_{2,i} \geq 0\},
\end{aligned}
\end{equation}
whose optimal value is denoted by $\tilde{q}(\cdot)$, is a convex relaxation of \eqref{o2}. In the optimization problem \eqref{sdp1}, $\mathcal{W}_{2,i} = X_iA_{cl,i}^T+A_{cl,i}X_i-X_iU^T-UX_i, \mathcal{S}_2 \triangleq \{X_i,\gamma_{1,i},\gamma_{2,i}\}$ and $U \in \mathbb{R}^{2n_x \times 2n_x}$ is given by the Cholesky decomposition: ${C_{r}^TC_{r}} = U^TU. \hfill \square$
\end{theorem}
In Theorem \ref{thm1}, we proposed a convex relaxation of \eqref{o2}, which is non-convex in the design variable $B_a$. Next, we show that for a given $\delta_i$, the value of the relaxed problem \eqref{sdp1} serves as an upper bound to the value of \eqref{o2}. 
\begin{lemma}\label{thm3}
Let the optimal tuple of \eqref{sdp1} be represented by $({\bar{X}_i,\bar{\gamma}_{1,i}},\bar{\gamma}_{2,i})$. Then the tuple $({\bar{P}_i\triangleq \bar{X}_{i}^{-1},\bar{\gamma}_{1,i}},\bar{\gamma}_{2,i})$ is a feasible solution to the optimization problem \eqref{o2}. Consequently, the optimal value of \eqref{sdp1} is greater than or equal to the optimal value of \eqref{o2}. $\hfill \square$
\end{lemma}

In this section, we proposed an SDP, convex in $B_a$ to determine the upper bound $\tilde{q}(B_a,\delta_i)$ for any given $B_a$ and sampled uncertainty $\delta_i$. The upper bound $\tilde{q}(\cdot)$ can act as a proxy for the impact $q(\cdot)$ and provide a certificate of the magnitude of the impact. In the next section, we relax the non-convex constraint $(C2)$
\begin{remark}\label{rem:minimal}
Assumption \ref{Ass:minimal} is necessary to prove Lemma \ref{thm0} using dissipative system theory (DST). When this assumption fails, there exists a sub-space of the closed-loop system that is uncontrollable/observable (UCO). For the exact treatment of such systems, the UCO subspace should be removed from the closed-loop system dynamics before formulating the Matrix Inequalities (MI) in \eqref{o2} using DST. However, if the closed-loop system is non-minimal, the MI in \eqref{dual} are only sufficient conditions for the constraints of \eqref{dual} to hold. Thus, when assumption \eqref{Ass:minimal} fails, \eqref{o2} represents an upper bound on the value of \eqref{dual}. $\hfill \triangleleft$
\end{remark}
\subsection{SDP relaxation of binary constraint}\label{sec:SDP_3}
Using the results of the previous section, to avoid the non-convex relation between $q(\cdot)$ and $B_a$ in \eqref{problem_1}, we replace $q(\cdot)$ by $\tilde{q}(\cdot)$ in \eqref{problem_1} and formulate \eqref{temp1}.
\begin{equation}\label{temp1}
\inf_{B_a} \left\{\text{CVaR}\{\tilde{q}(B_a,\delta)\}\Big| (C1),(C2)\right\}.
\end{equation}

The optimization problem \eqref{temp1} is non-convex since it involves SDP constraints with binary variables $(C2)$. As a first step toward relaxing $(C2)$, we reformulate \eqref{temp1} next.
\begin{lemma}\label{lem_relax_binary}
The optimization problems \eqref{temp1} and \eqref{temp2} are equivalent. 
\begin{equation}\label{temp2}
\begin{aligned}
\inf_{Z, z \in \mathbb{R}^{n_u}}\quad& \text{CVaR}\{\tilde{q}(\text{diag}(z),\delta)\}.\\
\text{s.t.}\;\;\quad&\begin{bmatrix}
    Z & z\\z^T & 1
\end{bmatrix} \succ 0,\; \sum_{i=1}^{n_u} z_i \geq n_u-n_w,\;(C10)\\
& \text{diag}(Z)=z,\;\text{rank}(Z) =1.
\end{aligned}
\end{equation}
\end{lemma}

In Lemma \ref{lem_relax_binary}, we reformulated \eqref{temp1} with binary constraints as \eqref{temp2}. However, this reformulation has rank constraints due to which \eqref{temp2} is still non-convex. To make the design problem convex, we remove the rank constraint.

\begin{corollary}\label{cor2}
A convex relaxation of \eqref{temp2} is given by
\begin{equation}\label{o5}
\inf_{Z, z}\left\{ \text{CVaR}\{\tilde{q}(\text{diag}(z),\delta)\}\Big|
(C10),\;\text{diag}(Z)=z. \right\}
\end{equation}
\end{corollary}

Corollary \ref{cor2} provides a method to relax $(C2)$ as an LMI constraint. There are many approaches in the literature to relax a binary variable constraint \cite[Table 1]{yuan2016binary}. However, we chose an LMI relaxation due to its simplicity. 

The result $z$ from \eqref{o5} will be integer instead of binary-valued. However, from Lemma \ref{lem_relax_binary}, we know that if the optimal $Z$ from \eqref{o5} has rank $1$, then the solution of \eqref{o5} is equal to the solution of \eqref{temp2}, and will be binary. For the general case, when the rank constraint is not satisfied, we provide a heuristic to convert the integers to binary variables later \cite{fischetti2010heuristics}. Next, we approximate the risk metric empirically. 
\subsection{Empirical approximation of CVaR}\label{sec:SDP_2}
The optimization problem \eqref{o5} is hard to solve since the CVaR operates over the set $\Omega$ which is a continuum (a similar observation was made in \cite{muller2018risk}). However, when we replace the uncertainty set $\Omega$ with, a sampled set with $N$ samples, the CVaR can be approximated by \cite[(9)]{rockafellar2000optimization}\\
\begin{strip}
\begin{equation}\label{sdp2}
 \begin{aligned}
\min_{\mathcal{S}_5}\quad&v+\frac{1}{N(1-\alpha)} \sum_{i=1}^N t_i\\
\textrm{s.t.}\quad&\begin{bmatrix}
-I & 0 & C_{p}X_i & 0\\ 0 & -\gamma_{1,i}I & I & 0\\ X_iC_{p}^T & I & \mathcal{W}_{2,i} & B_{cl,i}(z)\\0 & 0 & B_{cl,i}(z)^T & -\gamma_{2,i}I\end{bmatrix}\preceq0&\forall i\in\Omega_N &\quad(C5)\\
& t_i \geq \epsilon_r\gamma_{1,i} + \epsilon_a\gamma_{2,i} -v,& \forall i \in \Omega_N &\quad(C6)\\
& t_i \geq 0,& \forall i \in \Omega_N &\quad(C7)\\
& X_i \succ 0, & \forall i \in \Omega_N &\quad(C8)\;\\
& \gamma_{1,i} \geq 0, \gamma_{2,i} \geq 0,& \forall i \in \Omega_N &\quad(C9)\\
& \begin{bmatrix}
Z & z\\z^T & 1
\end{bmatrix} \succeq 0,\;n_u-n_w \leq \sum_{i=1}^{n_u} z_i & \; &\quad(C10)\\
& \mathcal{S}_5 \triangleq \{z,Z\} \cup \left\{\cup_{i=1}^N \left\{t_i, X_i, \gamma_{1,i},\gamma_{2,i}\right\}\right\}\\
& B_a = \text{diag}(z),\;\text{diag}(Z)=z, \Omega_N=\{1,\dots,N\}.
\end{aligned}
\end{equation}
\end{strip}

\begin{equation}\label{o3}
\text{CVaR}_{\alpha}\left\{\tilde{q}(\cdot,\delta)\right\} \approx\inf_v v + \frac{1}{1-\alpha} \frac{1}{N}\sum_{i=1}^N[\tilde{q}(\cdot,\delta_i)-v]^{+},
\end{equation}
where given $X \in \mathbb{R}, [X]^{+} \triangleq \max\{X,0\}$. Thus using \eqref{o3}, \eqref{o5} can be written as
\begin{equation}\label{o4}
\inf_{z,Z}\left\{ v+\frac{1}{N(1-\alpha)}\sum_{i=1}^N[\tilde{q}(\cdot)-v]^{+}\Bigg|\;
\begin{matrix}
    (C10),\\
    \text{diag}(Z)=z
\end{matrix}\right\}
\end{equation}
Now we briefly comment on the convergence of the empirical CVaR \eqref{o4} to the true CVaR \eqref{o5}. However, the proof of the following lemma is omitted as it is similar to the proof of \cite[Theorem 6]{muller2022risk}.

\begin{lemma}\label{lem_converge}
Let $\alpha$ represent the risk threshold. Given $N$ and $\alpha$, let $\tilde{r}_N$ represent the optimal value of \eqref{o4}, and let $\tilde{r}$ represent the optimal value of \eqref{o5}. Then it holds that $\lim_{N\to\infty} \tilde{r}_N \to \tilde{r}.$ $\hfill\;\square$
\end{lemma}

Lemma \ref{lem_converge} states that the empirical CVaR almost surely converges to the true CVaR in the large sample case. Now, we present a convex SDP to solve \eqref{o4} in Lemma \ref{thm2}.

\begin{lemma}\label{thm2}
Let us represent the optimal value of \eqref{sdp2} as $\underline{\gamma}$, and the optimal argument of $z \in \mathbb{R}^{n_u}$ from \eqref{sdp2} as $\underline{z}$.
Then an approximate binary solution to \eqref{o4} is given by 
\begin{equation}\label{eq:last}
    \bar{B}_a(i,i) =  \begin{cases} 
      0, & \text{if}\; \underline{z}_i \;\text{belongs to statistics of}\\
      \; & \text{order}\; 1, 2,\dots,\text{or}\;n_w\\
      1, & \text{otherwise}.\hfill \square
      \end{cases}
\end{equation}
\end{lemma}

The optimizer $z$ in \eqref{sdp2} is the diagonal of $B_a$. To represent the dependence of the constraint $(C5)$ (in \eqref{sdp2}) on $z$, the matrix $B_{cl,i}(z)$ (which is a function of $B_a$) is written as a function of $z$. And \eqref{eq:last} in Lemma \ref{thm2} is a heuristic to convert the decision variables ($z \in \mathbb{R}^{n_u}$) to binary variables.

In this article, to solve the security allocation problem via an SDP, we introduced some approximations. Next, we provide some discussions on the optimality of the solution obtained via these approximations. 
\begin{theorem}\label{prop_2}
    Let us represent the optimal solution obtained from \eqref{sdp2} as $\underline{z}$, the value of \eqref{sdp2} as $\underline{\gamma}$, the value of \eqref{temp1} as $\gamma^*$ (albeit unknown), the approximate solution obtained from \eqref{eq:last} as $\bar{B}_{a}$. Then the following statements are true.
    \begin{enumerate}[label=(\alph*)]
        \item (No loss of optimality) If rank$(\bar{Z})=1$, then $\bar{B}_{a}$ is {an} optimal solution to \eqref{temp1}.
        \item (When sub-optimal, characterizing a posteriori bound for the optimal risk $\gamma^*$) Let 
        \begin{equation}
            \bar{\gamma} \triangleq \inf_v \left\{v + \frac{1}{1-\alpha} \frac{1}{N}\sum_{i=1}^N[\tilde{q}(\bar{B}_{a},\delta_i)-v]^{+}\right\}
        \end{equation}
        where the value of $\tilde{q}(\bar{B}_{a},\delta_i)$ {is computed using the SDP \eqref{sdp1}}. Here $\bar{\gamma}$ is the CVaR under the sub-optimal attack matrix $\bar{B}_{a}$. Then, it holds that $\underline{\gamma} \leq \gamma^* \leq \bar{\gamma}.$
        \item (When sub-optimal, characterizing a posteriori upper bound for the optimality gap) Let us define the optimality gap as the difference between the true CVaR $\bar{\gamma}$, and the optimal CVaR $\gamma^*$. Then it holds that$$|\bar{\gamma}-\gamma^{*}| \leq |\bar{\gamma}-\underline{\gamma}|. \;\;\square$$
    \end{enumerate}
\end{theorem}
Theorem \ref{prop_2} states that the value of the optimal risk ($\gamma^*$) albeit unknown is bounded above and below by known values. It also provides an upper bound on the difference between the true risk ($\bar{\gamma}$) and the optimal risk ($\gamma^*$). In the next section, we discuss the solution to the allocation problem under different risk metrics.
\begin{remark}
[Actual risk incurred]
Let $\zeta= \inf_v v + \frac{1}{1-\alpha} \frac{1}{N}\sum_{i=1}^N[{q}(\bar{B}_{a},\delta_i)-v]^{+}$ where the value of ${q}(\cdot,\cdot)$ is determined from \eqref{o2}. Then the risk incurred by implementing the attack matrix $\bar{B}_{a}$ is $\zeta.\hfill\;\triangleleft$
\end{remark}
\section{Alternative risk measures}\label{sec:worst}
The previous section focussed on providing an (approximate) solution to the allocation problem \eqref{problem_1} which considered the risk metric CVaR. For the sake of comparison, we briefly study the allocation problem using two other measures of risk $(i)$ Worst case measure, and $(ii)$ nominal measure.
\subsection{Worst-case measure}\label{sec:WC}
For any random variable $X(\cdot,\delta),\delta \in \Omega$, the worst case measure is defined as $\sup_{\delta \in \Omega}X(\cdot,\delta)$: which represents the maximum loss that can occur. Then, under the worst-case measure, the allocation problem formulated in \eqref{eq:problem_0} becomes $$\arg\inf_{B_a} \left\{\sup_{\delta \in \Omega}\{q(B_a,\delta)\}\Big| (C1),(C2)\right\}$$

Similar to approximations in Section \ref{sec:SDP}, we first replace $q(\cdot)$ with $\tilde{q}(\cdot)$ to make the problem convex. Then we replace $\Omega$ with the sampled set. Then the design problem becomes 
\begin{equation}\label{wc_app}
\arg\inf_{B_a} \left\{\sup_{\delta_i, i \in \Omega_N}\{\tilde{q}(B_a,\delta_i)\}\Big| (C1),(C2)\right\}.
\end{equation}

Next, we propose an approximate solution to \eqref{wc_app} in Lemma \ref{lem_wc} using similar methods adopted in Lemma \ref{thm2}.
\begin{lemma}\label{lem_wc}
Let $z \in \mathbb{R}^{n_u}$. Let $z^*$ represent the optimal argument of $z$ from the SDP \eqref{sdp3}.
\begin{equation}\label{sdp3}
\begin{aligned}
\min_{\mathcal{S}_6} \quad & t\\
\textrm{s.t.} \quad &t \geq \epsilon_r\gamma_{1,i} + \epsilon_a\gamma_{2,i},\;\; \forall i \in \Omega_N\\
& (C5), (C8) - (C11)
\end{aligned}
\end{equation}
where $\mathcal{S}_6 = \{z,t\} \cup \left\{\cup_{i=1}^N \left\{X_i, \gamma_{1,i},\gamma_{2,i}\right\}\right\}, B_a = \text{diag}(z)$, and $Z=\text{diag}(z)$. Then an approximate binary solution to \eqref{wc_app} is given by \eqref{eq:last}.$\hfill \blacksquare$
\end{lemma}

\subsection{Nominal measure}\label{sec:NC}
Although we use risk measures for allocation in uncertain systems, it is logical to ask the question: \say{Is considering risk metrics necessary?}. To answer this question, we outline the allocation strategy when uncertainties are not considered. In other words, we allocate the security measures for the nominal system: $\inf_{B_a} \left\{q(B_a,\emptyset)\Big| (C1),(C2)\right\}.$ Then, similar to \eqref{wc_app}, we relax the allocation problem as 
\begin{equation}\label{nc_app}
\inf_{B_a} \left\{\tilde{q}(B_a,\emptyset)\Big| (C1),(C2)\right\}.
\end{equation}

Next, we propose an approximate solution to \eqref{nc_app} by a similar method adopted in Lemma \ref{lem_nc} whose proof is omitted since it is similar to the proof of Lemma \ref{lem_wc}.
\begin{lemma}\label{lem_nc}
Let $A_{cl}$, and $B_{cl}$ denote the nominal system matrices of \eqref{system:CL:uncertain}. And let $z \in \mathbb{R}^{n_u}$. Let us represent the optimal argument of $z$ from the SDP \eqref{sdp4} as $z^*$.
\begin{equation}\label{sdp4}
\begin{aligned}
\min_{\mathcal{S}_7} \quad & \epsilon_r\gamma_{1} + \epsilon_a\gamma_{2}\\
\textrm{s.t.} \quad& \begin{bmatrix}
-I & 0 & C_{p}X & 0\\ 0 & -\gamma_{1}I & I & 0\\ XC_{p}^T & I & \mathcal{W}_{2} & B_{cl}(z)\\0 & 0 & B_{cl}(z)^T & -\gamma_{2}I \end{bmatrix} \preceq 0\\
& \mathcal{W}_{2} = XA_{cl}^T+A_{cl}X-XU^T-UX\\
& (C10), (C11),\\
& \mathcal{S}_7 = \{z, X \succ 0, \gamma_{1} \geq 0,\gamma_{2} \geq 0\}.
\end{aligned}
\end{equation}
where $B_a = \text{diag}(z)$, and $Z=\text{diag}(z)$. Then an approximate binary solution to \eqref{nc_app} is given by \eqref{eq:last}. $\hfill \blacksquare$
\end{lemma}

In this section, we outlined the solution to the allocation problem under two other risk metrics. However, in the method that we propose to solve the allocation problem (in Lemma \ref{thm2}, Lemma \ref{lem_wc}, and Lemma \ref{lem_nc}), there are two sources of suboptimality. The first is the convex relaxation in formulating the convex upper bound $\tilde{q}(\cdot)$, and the second while relaxing the non-convex binary constraint ($C2$). 

In the next section, we present two algorithms: an algorithm that is computationally intensive but strictly optimal (exhaustive search), and a greedy algorithm that is polynomial in time but without any optimality guarantees. We also discuss the (de)merits of all three methods.
\section{Alternative search algorithms}\label{sec:exhaust:greedy}
In this section, we outline a method to determine the optimal solution of \eqref{problem_1}. Before this, we introduce the following notations. The set of all actuators is represented by $\mathcal{A}$, and for any finite set $\mathcal{Q}$, an element of $\mathcal{Q}$ is represented by $q$.
\subsection{Exhaustive search}
The exhaustive search algorithm first determines all possible subsets of $\mathcal{A}$ with maximum cardinality $n_w$. Then, it determines the CVaR when these various subsets of actuators are protected. Then the optimal solution to the allocation problem is the set of actuators that yields the minimum CVaR. We outline an exhaustive search in Algorithm \ref{algo1}, where $g^*$ represents the optimal set of protected actuators.

In Algorithm \ref{algo1}, if the CVaR is determined using $q(\cdot,\delta_i)$ in \eqref{o2}, the result of the algorithm is optimal. The result of Algorithm \ref{algo1} can then be then used to compare how the approximation in formulating $\tilde{q}$ affects the solutions in \eqref{sdp2}. However, if the CVaR is determined using $\tilde{q}(\cdot,\delta_i)$ in \eqref{sdp1}, the algorithm is sub-optimal. 

The time complexity of the exhaustive search is very high since the algorithm searches over all possible choices of actuators. Next, we discuss a greedy algorithm that is polynomial in time but provides a sub-optimal solution. 
\begin{algorithm}
\SetAlgoLined
\caption{Exhaustive search to solve \eqref{problem_1}}
\textbf{Initialization}: $\alpha,\Omega_N, \mathcal{A},n_w$ and an empty list $\gamma$\\
\textbf{Step 1:} Determine $\mathcal{G}$ as the set of all subsets of $\mathcal{A}$ with cardinality $n_w$.\\
\textbf{Step 2:} \\
\ForAll{$g \in \mathcal{G}$}{
Set $B_a(i,i) = 0$ if $i \in g$ and $1$ otherwise.\\
Determine the CVaR$_{\alpha}\{{q}(B_a,\delta)\}$ \eqref{o3} with this new $B_a$.\\
Append $\{\text{CVaR}_{\alpha}\{{q}(B_a,\delta)\},g\}$ to the list $\gamma$\\
  }
\textbf{Step 3} Determine $\gamma^*=\min_{j} \text{CVaR}^{[j]}_{\alpha}\{{q}(B_a,\delta)\}$ and the respective $g^*=g^{[j^*]}$\\
\KwResult{$g^*$ $\hfill \triangleleft$} 
\label{algo1}
\end{algorithm}
\subsection{Greedy search}
The greedy algorithm first chooses one actuator to be protected which minimizes the CVaR. Let this actuator be the first actuator $a_1$. Now with $a_1$ being protected, the algorithm searches for one more actuator to be protected such that the actuator pair $\{a_1\} \cup \{a_{l}\}, l \in \{2,\dots,n_a\}$ minimizes the CVaR. Let this actuator pair be $\{a_1,a_6\}$. In this way, the greedy algorithm continues searching for one actuator to protect at a time which minimizes the CVaR until the number of protected actuators is $n_w$. This greedy algorithm is depicted in Algorithm \ref{algo2}. 

\begin{algorithm}
\caption{Greedy search to solve \eqref{problem_1}}
\textbf{Initialization}: $\alpha, \Omega_N, \mathcal{A},n_w$, and empty lists $\gamma,\mathcal{W}$\\
\For{$j=1:n_w$}{
Clear the list $\gamma$\\
  \For{$i=1:n_u$}{
 Set $
    B_a(s,s) = \begin{cases} 
      0, & \text{if}\;s\in \mathcal{W},\;\text{or}\; s=i \\
      1, & \text{otherwise}.
      \end{cases}$\\
Determine the CVaR$_{\alpha}{\tilde{q}(B_a,\delta)}$ \eqref{o3} using the new $B_a$.\\
Append $\gamma$ with CVaR$_{\alpha}{\tilde{q}(B_a,\delta)}$\\
  }
Determine $\gamma^*= \min_{k=\{1,2,\dots,n_u\}} \gamma^{[k]}$ and the respective $k^*$ \\
Append $k^*$ to $\mathcal{W}$.\\
}
\KwResult{$\mathcal{W}$ $\hfill \triangleleft$}
\label{algo2}
\end{algorithm}

In Algorithm \ref{algo2}, the result $\mathcal{W}$ represents the sub-optimal set of actuators to be protected. The result is suboptimal since the algorithm does not search over all sets of possible actuators. The greedy algorithm is included in this article for comparison of performance. Also, if the submodularity and non-increasing property of CVaR($q(\cdot)$) is proven, then the greedy algorithm can give certain performance guarantees \cite{wolsey1982analysis}: which is left for future work. 

So far, we discussed three methods to (approximately) solve \eqref{problem_1}. Our proposed SDP method \eqref{sdp2} is an approximate solution and has polynomial time complexity in the worst case. The exhaustive search in Algorithm \ref{algo1} provides the optimal solution but has combinatorial complexity. Finally, the greedy algorithm is also polynomial in time complexity but provides a sub-optimal solution. However, as mentioned before, the greedy algorithm has some scope for future work. Next, we compare the methods through a numerical example.

\begin{remark}
The exhaustive and greedy search algorithms can also be used with other risk metrics. For instance, instead of CVaR, we can determine the worst case or the nominal measure of $\tilde{q}(\cdot)$ in Algorithm \ref{algo1} and Algorithm \ref{algo2}. However, we do not detail this due to lack of space. $\hfill \triangleleft$
\end{remark}
\section{Numerical example}\label{sec:NE}
\begin{figure}
    \centering
    \includegraphics[width=18.5pc]{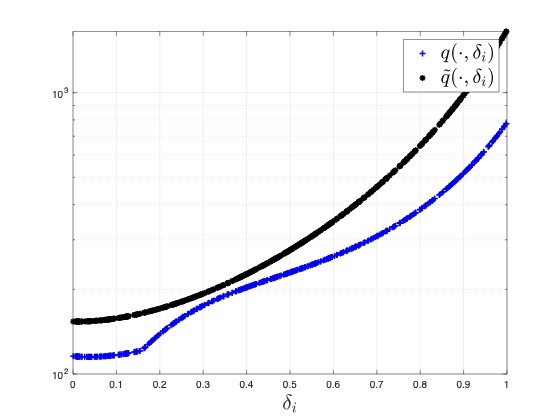}
    \caption{The values of $q(B_a=I_4,\delta_i)$ and $\tilde{q}(B_a=I_4,\delta_i)$ across different $\delta_i$, obtained by solving \eqref{o2} and \eqref{sdp1} respectively.}
    \label{fig2}
\end{figure}

The effectiveness of the method discussed in Lemmas \ref{thm2} is illustrated through numerical examples in this section. Consider the system in \eqref{P}-\eqref{D} with matrices given in \eqref{matrix_NE} and $B=B_a=I_4.$
We set $\epsilon_r=1, \epsilon_a=300$, and $N=500$. We sample $\Omega$ according to sample distribution. Then, we determine the value of $q(B_a,\delta_i)$ using \eqref{o2} and $\tilde{q}(B_a,\delta_i)$ using \eqref{sdp1} which are plotted in Figure \ref{fig2}.

In line with Remark \ref{rem_1}, the value of $q(\cdot,\delta)$ is bounded for all uncertainties $\delta \in \Omega$. To recall, the value of $q(\cdot,\delta)$ (attack impact) is bounded since the attack energy is bounded. 
Also, in line with Lemma \ref{thm3}, ${q}(\cdot)$ is upper bounded by $\tilde{q}(\cdot)$. 

The rest of this section is organized as follows. In Section \ref{sec_NE_0}, we compare the metric \eqref{o2} to other security metrics in the literature. In Section \ref{sec_NE_1} we compare the results to the allocation problem when using CVaR and the nominal measure, whereas in Section \ref{sec_NE_2} we compare CVaR against the worst-case measure. In section \ref{sec_NE_3}, we compare the different search algorithms. Finally, in Section \ref{sec_NE_4}, we compare the solution from \eqref{sdp2} to the optimal solution.
\subsection{Comparison with other metrics}\label{sec_NE_0}
Following the discussion in section \ref{sec:discuss}, to compare our metric \eqref{eqq4} to other security metrics, we proceed as follows. We set $B^T=\begin{bmatrix}1 & 0 & 0 & 0\end{bmatrix}$, and $B_a=1$. Then we determine the value of $q(\cdot,\emptyset)$ by solving \eqref{o2} (equivalent to \eqref{eqq4}) when $\epsilon_a=10^6$ and $\epsilon_r=1$. This makes the constraint on the attack energy inactive making $q(\cdot)$ the OOG. We found this value to be $34.45$. Next, we determine the true OOG by solving \cite[(6.18)]{ferrari2021safety} and these values match. 

We set $\epsilon_r = 10^6$ and $\epsilon_a=1$. This makes the constraint on the detection output inactive, making $q(\cdot)$ the $H_{\infty}$ metric. We found the value of $q(\cdot)$ to be $0.62$. We also determine the value of the $H_{\infty}$ metric by solving the LMI in \cite{hilhorst2014extended} and these values match. Thus we numerically depict the relation between \eqref{eqq4} and other metrics.
\begin{figure}
    \centering
    \includegraphics[width=18.5pc]{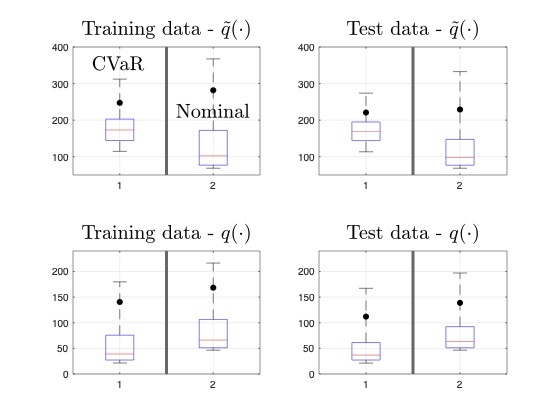}
    \caption[]{The box plots with $N=100$ in the top (bottom) depicts the value of the attack impact $q(\cdot)$ (the impact proxy $\tilde{q}(\cdot)$) when the protected actuators are obtained from optimizing the CVaR ($A_2$ and $A_4$) in \eqref{sdp2} and the nominal measure ($A_1$ and $A_4$) in \eqref{sdp4}. The plots on the left (right) represent values obtained from training (test) data. Here training data represents the data points (of uncertainty) used in the optimization problem, and test data represents new data points (of uncertainty). On each box, the central mark indicates the median, and the bottom and top edges of the box indicate the $25^{th}$ and $75^{th}$ percentiles, respectively, and the black dot represents the CVaR$_{0.8}$ of the data points. The whiskers extend to the most extreme data points.}
    \label{fig3}
\end{figure}
\subsection{Comparison with nominal measure.}\label{sec_NE_1}
Next we set $N=100, \alpha=0.8, \epsilon_r=1,$ and $\epsilon_a=300$. {For the sake of comparison, we determine the CVaR$_{0.8}(\tilde{q}(\cdot))$ when $n_w=0$ (no protection) as $2813.6$.} Next, we allocate the security measure that minimizes CVaR$_{0.8}(\tilde{q}(\cdot))$ by solving the optimization problem \eqref{sdp2} and obtain the actuators to be protected as $A_2$ and $A_4$ (here $A_i,i\in\{1,\dots,n_u\}$ represents the $i^{\text{th}}$ actuator). 

To depict the effectiveness of using a risk metric, we solve the allocation problem which minimizes $\tilde{q}(\cdot,\emptyset)$, i.e., using the nominal measure, by solving \eqref{sdp4}. We obtain the actuators to be protected as $\{A_1, A_4\}$.  

To visualize the effectiveness of the used metric, in Fig \ref{fig3}, we plot the value of the attack impact $q(\cdot)$, the impact proxy $\tilde{q}(\cdot)$ when the protected actuators are $\{A_2, A_4\}$, and $\{A_1, A_4\}$ respectively. Now some remarks are in order.

Firstly, as expected, we see that using the risk metric instead of the nominal measure reduces the CVaR (the black dots in Figure \ref{fig3}) across training and test data, and across $q(\cdot)$ and $\tilde{q}(\cdot)$. Secondly, using a risk metric minimizes the worst-case impact and the impact proxy (the top whiskers of the box plots in Fig \ref{fig3}). Thirdly, although the median of the impact proxy (the red horizontal lines in Figure \ref{fig3}) is higher when using the risk metric, the median of the actual impact $q(\cdot)$ is lower. Finally, we see that the $25^{th}$ percentile of the impact $q(\cdot)$ is lower when using the risk metric. 

\begin{figure}
    \centering
    \includegraphics[width=18.5pc]{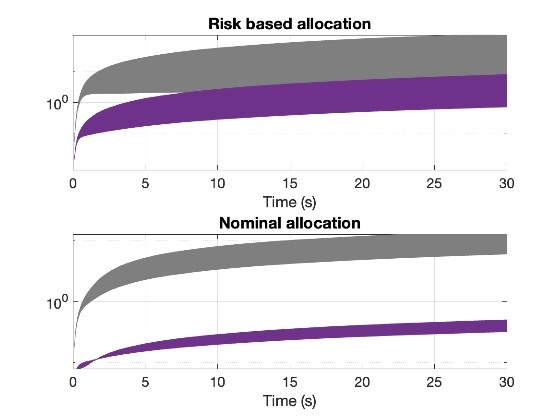}
    \caption{Performance energy (grey) and detection energy (violet) for $N=500$ different realizations of uncertainty, under CVaR-based allocation strategy (top), and the nominal allocation strategy (bottom).}
    \label{figa}
\end{figure}

Next we consider a step attack signal $a(t)=\textbf{1},\; t\geq 0$.
Under the step attack, the performance energies under $N=500$ different realizations of the uncertainty are shown in Figure \ref{figa}. The performance energy when the allocation is done by optimizing the CVaR is depicted at the top of Figure \ref{figa}, and the nominal measure is depicted at the bottom of Figure \ref{figa}. As mentioned before, the objective of the allocation problem is to minimize the performance loss under attacks. From Figure \ref{figa} we see that the worst-case performance loss is the same (approximately) under the different allocation strategies. However, under the CVaR-based allocation, the best-case performance loss is low, thus depicting an advantage.

The detection energies are depicted in violet colour in Figure \ref{figa}. As mentioned before, the objective of the allocation problem is to maximize the detection output energy and raise an alarm when $||y_r||_{\ell_2}^2 > \epsilon_r$. When $\epsilon_r=1$, under the nominal allocation strategy, we can see from Figure \ref{figa} that the alarm will never be raised, thus depicting a poor performance.  In other words, for attack detection, $\epsilon_r$ should be as low as $0.1$ which can be impractical in the presence of noise. However, under the CVaR-based allocation strategy, the attack is detected when $\epsilon_r=1$. Thus, our method can help to detect attacks better. The high performance deterioration under attack may be prevented by timely switching to a fault-tolerant controller when the attack is detected.

\subsection{Comparison with worst-case measure.}\label{sec_NE_2}
\begin{figure}
    \centering
    \includegraphics[width=18.5pc]{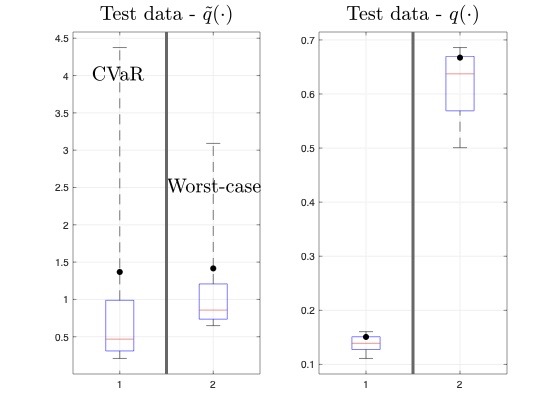}
    \caption{The box plots with $N=500$ in the left (right) depicts the value of the impact proxy $\tilde{q}(\cdot)$ (attack impact ${q}(\cdot)$) when the protected actuators are obtained from optimizing the CVaR ($\{A_1, A_2, A_3\}$) in \eqref{sdp2} and the worst-case ($\{A_1, A_2, A_6\}$) in \eqref{sdp3}. On each box, the black dot represents the CVaR$_{0.5}$ of the data points.}
    \label{fig4}
\end{figure}

For this comparison, we now consider a distributed NCS, consisting of agents with single integrator dynamics as described in \cite{nguyen2022zero}. The operator is uncertain about the edge weights of the undirected graph. Each agent has a wireless control loop that is prone to attack. The system matrices of the NCS (derived similar to \cite[(6)]{nguyen2022zero}) are $A_{cl}^{\Delta}=$
\begin{equation}
A_{cl}^{\Delta}=\begin{bmatrix}\label{ex_robot}
\delta-32 & 4 & 0 & 3 & 0 & 5+\delta\\
4 & -37 & 3 & 4 & 4 & 0\\
0  & 3 & -29 & 2 & 0 & 0\\
3 & 4 & 2 & -33 & 3 & 0\\
0 & 4 & 0 & 3 & -28 & 1\\
5+\delta & 0 & 0 & 0 & 1 & \delta-24\\
\end{bmatrix}
\end{equation}
where $\delta \in \Omega \triangleq [-1,\;0], B_{cl}=I_6,$ and $\begin{bmatrix}
C_p\\ \hline C_r
\end{bmatrix} \triangleq
\begin{bmatrix}
0 & 0 & 1 & 0 & 0 &0\\
\hline
0 & 0 & 0 & 0 &0 & 1
\end{bmatrix}$. Here $\delta$ represents the uncertainty in the edge weights of the NCS. We set $N=1000, n_w=3, \alpha=0.5$. We are now interested in allocating the security measure which minimizes the CVaR$_{0.5}(\tilde{q}(\cdot))$. To this end, we solve the optimization problem \eqref{sdp2} and obtain the actuators to be protected as $\{A_1, A_2, A_3\}$.  For comparison, we solve the allocation problem that minimizes the worst-case impact \eqref{sdp3}, and we obtain the actuators to be protected as $\{A_1, A_2, A_6\}$.
To visualize the effectiveness of the used metric, in Fig \ref{fig4}, we plot the values of the attack impact $q(\cdot)$ and the impact proxy $\tilde{q}(\cdot)$ for some test data when protected actuators are $\{A_1, A_2, A_3\}$, and $\{A_1, A_2, A_6\}$, respectively. 

Firstly, as expected, we see that using CVaR as a risk metric reduces the CVaR of $\tilde{q}(\cdot)$ (black dot in Figure \ref{fig4}). Secondly, using CVaR causes the worst-case impact (top whiskers of ${q}$) to be low. Finally, using the CVaR as a risk metric reduces the median (red horizontal line in the box plot), and the $25^{th}$ percentile across $q(\cdot)$ and $\tilde{q}(\cdot)$.
\subsection{Comparison with other search algorithms}\label{sec_NE_3}
Now we have shown the effectiveness of using CVaR as a risk metric. Next, we show the effectiveness of the proposed allocation method. To this end, we first solve the allocation problem through an exhaustive search. That is, we consider the system matrices in \eqref{ex_robot} and solve the allocation problem which minimizes the CVaR$_{0.3}$ by an exhaustive search as in Algorithm \ref{algo1}. Similarly, we also solve the problem using greedy search in Algorithm \ref{algo2}. We observe that the results are the same as obtained by using our SDP \eqref{sdp2}: that is, we obtain that the protected actuators are $\{A_1, A_2, A_3\}$. However, the time taken to obtain these results are significantly different and are given in TABLE \ref{t1}. The results are tabulated when $N=100$ and $N=200$. We can see that the computational time for the convex SDP that we propose in this article is at least $40$ times faster than the other two methods, thereby depicting its efficacy. 
\subsection{Comparison to the optimal solution}\label{sec_NE_4}
Next, we discuss the loss of optimality in the proposed SDP \eqref{sdp2} due to the approximation in formulating $\tilde{q}$. We compare the solution obtained from \eqref{sdp2} to the solution obtained from Algorithm \ref{algo1} when $q(\cdot)$ from \eqref{o2} is used to determine the CVaR. Recall that when $q(\cdot)$ from \eqref{o2} is used in Algorithm \ref{algo1}, it provides the optimal solution.

As we already know, the solution from \eqref{sdp2} is $\{A_1, A_2, A_3\}$. We obtain the optimal solution from Algorithm \ref{algo1} to be $\{A_2,A_3,A_6\}$ when $q(\cdot)$ is used. Thus, we can see that there is a loss of optimality here. However, we report that the difference in the CVaR between these two solutions in the test data is only $0.02$ which is negligible. 
\begin{table}
\centering
\begin{tabular}{|| c | c | c ||}
 \hline
 Method & $N=100$ & $N=200$ \\
 \hline 
 SDP \eqref{sdp2} & $3.77$ \mbox{sec} & $7.77$ \mbox{sec}\\
 \hline
Algorithm \ref{algo1} (Exhaustive search) & $262.30$ \mbox{sec} & $523.29$ \mbox{sec} \\
 \hline
 Algorithm \ref{algo2} (Greedy search) & $243.26$ \mbox{sec} & $469.19$ \mbox{sec}\\
 \hline
\end{tabular}
\caption{Comparison of results}
\label{t1}
\end{table}
\section{Conclusions}\label{sec:Conclusion}
This article considered the problem of security measure allocation when the actuators of an uncertain NCS are under attack. The CVaR was used to formulate the risk associated with the attack impact. The allocation problem was observed to be hard to solve since it involves SDP constraints with binary decision variables. Thus we use Young's relation to formulate a relaxed convex SDP. We also briefly compare our algorithm across different risk metrics and different search algorithms: discussing its merits and demerits. The efficacy of our proposed approach is discussed through numerical examples. Future works include providing any performance guarantees on the proposed approach. 
\renewcommand{\thesection}{A.\arabic{section}}
\setcounter{section}{0}  
\renewcommand{\thedefinition}{A.\arabic{section}.\arabic{definition}}
\setcounter{definition}{0}
\renewcommand{\thetheorem}{A.\arabic{section}.\arabic{theorem}}
\setcounter{theorem}{0}
\renewcommand{\theassumption}{A.\arabic{section}.\arabic{assumption}}
\setcounter{assumption}{0}
\renewcommand{\theremark}{A.\arabic{section}.\arabic{remark}}
\setcounter{remark}{0}
\renewcommand{\theprop}{A.\arabic{section}.\arabic{prop}}
\setcounter{prop}{0}
\section*{Appendix}
\section{Proof of Lemma \ref{lem0}}
\begin{proof}
Consider the constraint $\Vert a_i \Vert_{\mathcal{L}_2}^2 \leq \epsilon_a$ in \eqref{eqq4}. We know that $\Vert a_i \Vert_{\mathcal{L}_2}^2 \leq \epsilon_a \implies \lim_{t \to \infty}a_i(t)=0$. Since the closed loop system is stable, $\lim_{t \to \infty}a_i(t)=0 \implies \lim_{t \to \infty}x_i(t)\triangleq x_i(\infty) =0$. Then, for a given $\delta_i$, $q(B_a,\delta_i)$ in \eqref{eqq4} can be reformulated using the hypergraph formulation as \eqref{o1}.
\begin{equation}\label{o1}
\sup_{\upsilon, a \in \mathcal{L}_{2e}} \left\{  \upsilon \; \Bigg|\; \begin{aligned}
& ||y_{p,i}||_{\mathcal{L}_{2}}^2 \geq \upsilon & ||y_{r,i}||_{\mathcal{L}_{2}}^2 \leq \epsilon_r\\
& ||a_{i}||_{\mathcal{L}_{2}}^2 \leq \epsilon_a & x_i(\infty)=0
\end{aligned}\right\}
\end{equation}
Note that \eqref{o1} is similar to \cite[(22)]{anand2021risk}. Then, following the proof of \cite[Theorem 4.4]{anand2021risk}, \eqref{o1} can be rewritten as \cite[(51)]{anand2021risk} which concludes the proof. 
\end{proof}
\section{Proof of Lemma \ref{thm0}}
Before we present the proof, we present an intermediate result which helps in constructing the proof of Lemma \ref{thm0}.
\begin{prop}[\cite{trentelman1991dissipation}]\label{prop1}
Consider a CT system $\Sigma \triangleq$ $ (A, B, C, D)$ which is controllable and observable with supply rate $s[\cdot] = ||y_1(t)||_2^2 - ||y_2(t)||_2^2 + ||u(t)||_2^2 $. Let $y_i(t) = C_i x(t) + D_i u(t),\; i=\{1,2\}$. Then the following statements are equivalent:
\begin{enumerate}
    \item For all trajectories of the system, for $T>0$ and $x[0]=0$, we have $\int_{0}^{T} s[x(t),u(t)]dt \geq 0$.
    \item There exists a symmetric $P\succeq 0$ such that \eqref{LMI1} holds.
    \begin{equation}\label{LMI1}
    \begin{bmatrix} A^T+PA & PB \\ B^TP & 0 \end{bmatrix} + R \preceq 0, 
    \end{equation}  
    $R \triangleq \begin{bmatrix}
    C_1^T\\D_1^T
    \end{bmatrix}\begin{bmatrix}
    C_1 & D_1
    \end{bmatrix}-\begin{bmatrix}
    C_2^T\\D_2^T
    \end{bmatrix}\begin{bmatrix}
    C_2 & D_2
    \end{bmatrix}-
    \begin{bmatrix}
    0 & 0\\0 & I
    \end{bmatrix} \hfill\square$
\end{enumerate}
\end{prop}
\begin{remark}[\cite{goodwin2014adaptive}]\label{rem:definite} 
Suppose that $(i)$ $\Sigma$ is minimal, and $(ii)$ for all $0 \neq y=\begin{bmatrix} y_1^T y_2^T\end{bmatrix}^T, \exists u$ such that $s[\cdot]<0$, then $\Sigma$ is dissipative iff $\exists P\succ 0$ such that \eqref{LMI1} holds.$\hfill \triangleleft$
\end{remark}
Next, we present the proof of Lemma \ref{thm0}. 
\begin{proof}
Let us define a (supply rate) function $s[\cdot] \triangleq -\Vert y_p(t) \Vert_{\mathcal{L}_2}^2 + \gamma_1 \Vert y_r(t)\Vert_{\mathcal{L}_2}^2+\gamma_2 \Vert a(t) \Vert_{\mathcal{L}_2}^2$ which is also the constraint of the optimization problem \eqref{dual}. Recall that the signals ($y_p,y_r,u$) obey the condition of Proposition \ref{prop1}: they originate from a system that is controllable and observable (Assumption \ref{Ass:minimal}). Then using \eqref{LMI1}, the constraint of \eqref{dual} can be replaced by \eqref{o2}. It only remains to show that $P \succ 0$. 

It holds that $P \succ 0$ if the conditions of Remark \ref{rem:definite} hold which we show next. Condition $(i)$ holds from Assumption \ref{Ass:minimal}. And for any non-zero $y$, $\gamma_2$ in $s[\cdot]$ can be increased arbitrarily such that $s[\cdot]<0$. Thus, the conditions of Remark \ref{rem:definite} hold which concludes the proof.
\end{proof}
\section{Proof of Theorem \ref{thm1}}
\begin{proof}
Applying Schur complement, $(C3_i)$ in \eqref{o2} becomes
\begin{equation}
\mathcal{W}_3 \triangleq \begin{bmatrix}
-I&C_{p}&0\\C_{p}^T&A_{cl,i}^TP_i+P_iA_{cl,i}-\gamma_{1,i}U^TU&P_iB_{cl,i}\\0&B_{cl,i}^TP_i&-\gamma_{2,i}I
\end{bmatrix}\preceq0
\end{equation}
We now apply congruence transformation \cite[Section 2.2]{caverly2019lmi} which states that the matrix inequality $\mathcal{W}_3 \preceq 0$ is satisfied if and only if $Z\mathcal{W}_3Z^T \preceq 0$ where $\text{rank}(Z)=n$. We pick $Z=\text{diag}(I,P_i^{-1},I)$. Then the first constraint of \eqref{o2} becomes 
\begin{equation}\label{t3}
\begin{bmatrix}
-I & C_{p}X_i & 0\\ X_iC_{p}^T & X_iA_{cl,i}^T+ A_{cl,i}X_i -\mathcal{W}_4 & B_{cl,i}\\0 & B_{cl,i}^T & -\gamma_{2,i}I
\end{bmatrix} \preceq 0
\end{equation}
where $\mathcal{W}_4 \triangleq \gamma_{1,i} X_iU^TUX_i$ and $X_i=P_i^{-1}$. Up to now, we have shown that \eqref{o2} (or equivalently \eqref{t3}) is convex in $B_a$ (since $B_{cl,i}$ is linear in $B_a$) except $\mathcal{W}_4$. We next approximate 
\begin{equation}\label{young}
\mathcal{W}_4 \succeq_{1} UX_i + X_iU-\gamma_{1,i}^{-1}I \triangleq \tilde{\mathcal{W}}_4,    
\end{equation}
where the inequality ${1}$ is from Young relation which is given by $\gamma^{-1}G^TG \succeq G + G^T \gamma I$ for any given matrix $G$ and $\gamma\geq 0$ \cite[Section 2.4.3]{caverly2019lmi}. We now relax the constraint \eqref{t3} by replacing ${\mathcal{W}}_4$ by $\tilde{\mathcal{W}}_4$. Then taking the Schur complement of the relaxed constraint concludes the proof.
\end{proof}
\section{Proof of Lemma \ref{thm3}}
\begin{proof}
The optimal tuple for \eqref{sdp1} is represented by $({\bar{X}_i,\bar{\gamma}_{1,i}},\bar{\gamma}_{2,i})$, by applying Schur complement to its first constraint, we get 
\begin{equation}\label{p1}
    Q + \begin{bmatrix}
    -\bar{X}_iU-U\bar{X}_i+\gamma_{1,i}^{-1}I & 0\\0 & 0
    \end{bmatrix}\preceq 0.
\end{equation}
where $Q \triangleq \begin{bmatrix}
    \bar{X}_iA_{cl,i}^T+A_{cl,i}\bar{X}_i+\bar{X}_iC_p^TC_p\bar{X}_i& B\\B^T & -\gamma_{2,i}I
    \end{bmatrix}$. Then by using \eqref{young}, \eqref{p1} becomes 
\begin{equation}\label{p2}
    Q+\begin{bmatrix}
    \gamma_{1,i}\bar{X}_iU^TU\bar{X}_i & 0\\0 & 0
    \end{bmatrix} \preceq 0
\end{equation}
We apply congruence transformation with $Z=\text{diag}(\bar{P}_i,I)$. Then \eqref{p2} is equivalent to 
\begin{equation}\label{p3}
    \begin{bmatrix}
    A_{cl,i}\bar{P}_i+\bar{P}_iA_{cl,i}+C_p^TC_p-\gamma_{1,i}C_r^TC_r & \bar{P}_iB_{cl}\\B_{cl,i}^T\bar{P}_{i} & -\gamma_{2,i}I
    \end{bmatrix} \preceq 0
\end{equation}
which is the constraint of \eqref{o2}. This concludes the first part of the proof. We prove the second proof by contradiction. For a given uncertainty $\delta_i$, let the optimal tuple of \eqref{sdp1} be $(\cdot,\gamma_{1R},\gamma_{2R})$. Similarly, let the optimal tuple of \eqref{o2} be  $(\cdot,\gamma_{1O},\gamma_{2O})$. Let us assume that $\gamma_R = \epsilon_1\gamma_{1R}+\epsilon_2\gamma_{2R} < \epsilon_1\gamma_{1O}+\epsilon_2\gamma_{2O} = \gamma_O$. We know from Theorem \ref{thm3} that every feasible tuple of \eqref{sdp1} is a feasible tuple of \eqref{o2}. Then $\gamma_{1R},\gamma_{2R}$ is a feasible solution to \eqref{o2} which yields a lower value to \eqref{o2}. However, this contradicts the assumption and concludes the proof.
\end{proof}
\section{Proof of Lemma \ref{lem_relax_binary}}
\begin{proof}
Let $z$ be the diagonal elements of $B_a$ (recall that only the diagonal elements of $B_a$ are the design variables). Now we show that when the constraints of \eqref{temp2} are satisfied, the variable $z$ is binary. Using Schur complement, $(C10)$ can be rewritten as $Z-zz^T\geq0$. And since the $\text{rank}(Z) =1$, we can conclude $\mathcal{Z}\triangleq Z-zz^T=0$. Let us consider the diagonal elements of the matrix $\mathcal{Z}$, which yields $z_i(1-z_i)=0$ whose solutions are $z_i=\{0,1\}$. This concludes the proof. 
\end{proof}

\section{Proof of Lemma \ref{thm2}}
\begin{proof}
Consider the objective function in \eqref{o4}. Given $\delta_i$, let $\tilde{q}(B_a,\delta_i)-v \triangleq t_i$. Then the projection of $t_i$ on the positive real axis is achieved by the constraints ($C6$) and ($C7$). Then the value of $\tilde{q}(B_a,\delta_i)$ is given by solving the optimization problem \eqref{sdp1}. Thus $\tilde{q}(B_a,\delta_i)$ is replaced by the objective function of \eqref{sdp1} and the corresponding constraint ($C5$) is included. The constraint ($C1$) is re-written as ($C11$). Using Lemma \ref{lem_relax_binary} ($C2$) is relaxed as ($C10$). The optimal argument $z^*$ of \eqref{sdp2} will not be binary but integers. To this end, let $\mathcal{K}$ denote a set that contains the $n_w$ least elements in the value of $z^*$. Then, the actuator channel $i$ is protected if $z_i$ belongs $\mathcal{K}$. This concludes the proof.
\end{proof}
\section{Proof of Theorem \ref{prop_2}}
\textit{Proof of (a):} The optimization problem \eqref{sdp2} was formulated by removing the rank constraint rank$(Z)=1$ from \eqref{temp2}. However, if the rank constraint is satisfied implicitly, the solution $\bar{B}_a$ is optimal. This concludes the proof of (a). \\[0.1cm]
\textit{Proof of (b):} Let us consider the optimization problem \eqref{temp1} whose value is $\gamma^*$. The value of $\gamma^*$ is the same as the value of \eqref{temp2} since they are equivalent. In the optimization problem \eqref{sdp2}, we removed the rank constraint. Since \eqref{sdp2} is a minimization problem, its value will be lower than \eqref{temp2}. Thus it holds that $\underline{\gamma} \leq \gamma^*.$ Since the approximate solution $\bar{B}_{a}$ obtained from \eqref{eq:last} is sub-optimal, but feasible to the optimization problem \eqref{temp1}, the corresponding risk: $\bar{\gamma} = \text{CVaR}_{\alpha}(\tilde{q}(\bar{B}_{a},\cdot))$ will be higher than the true risk. Thus $\bar{\gamma}$ acts as an upper bound for the true risk $\gamma^*$. Then, the following holds $ \gamma^* \leq \bar{\gamma}.$ Combining the above two arguments concludes the proof of (b). \\[0.1cm]
\textit{Proof of (c):} The proof follows directly from (b), and the fact that the quantities $(\bar{\gamma}-\gamma^{*})$ and $\bar{\gamma}-\underline{\gamma}$ are positive. This concludes the proof. $\hfill \blacksquare$
\section{Proof of Lemma \ref{lem_wc}}
\begin{proof}
Using the hyper-graph formulation, and the SDP \eqref{sdp1}, the objective function in \eqref{wc_app}: $\sup \tilde{q}(\cdot)$, can be re-written as $t \geq \epsilon_r\gamma_{1,i} + \epsilon_a\gamma_{2,i},\forall i \in \Omega_{N}$. The corresponding constraint ($C4$) is included. The constraint ($C1$) is re-written as ($C11$). And an SDP relaxation of the constraint ($C2$) is formulated using ($C10$). The optimal argument $z^*$ of \eqref{sdp2} will not be binary but integers. To this end, let $\mathcal{K}$ denote a set that contains the $n_w$ least elements in the value of $z^*$. Then, actuator $i$ is protected if $z_i$ belongs $\mathcal{K}$. This concludes the proof.
\end{proof}

\bibliographystyle{ieeetr}
\bibliography{OJCSYS_template} 

\begin{thebibliography}{10}

\bibitem{sandberg2022secure}
H.~Sandberg, V.~Gupta, and K.~H. Johansson, ``Secure networked control
  systems,'' {\em Annual Review of Control, Robotics, and Autonomous Systems},
  vol.~5, pp.~445--464, 2022.

\bibitem{dibaji2019systems}
S.~M. Dibaji, M.~Pirani, D.~B. Flamholz, A.~M. Annaswamy, K.~H. Johansson, and
  A.~Chakrabortty, ``A systems and control perspective of cps security,'' {\em
  Annual reviews in control}, vol.~47, pp.~394--411, 2019.

\bibitem{chong2019tutorial}
M.~S. Chong, H.~Sandberg, and A.~M. Teixeira, ``A tutorial introduction to
  security and privacy for cyber-physical systems,'' in {\em 2019 18th European
  Control Conference (ECC)}, pp.~968--978, IEEE, 2019.

\bibitem{ferrari2021safety}
R.~M. Ferrari and A.~M. Teixeira, {\em Safety, Security and Privacy for
  Cyber-Physical Systems}.
\newblock Springer, 2021.

\bibitem{giraldo2018survey}
J.~Giraldo, D.~Urbina, A.~Cardenas, J.~Valente, M.~Faisal, J.~Ruths, N.~O.
  Tippenhauer, H.~Sandberg, and R.~Candell, ``A survey of physics-based attack
  detection in cyber-physical systems,'' {\em ACM Computing Surveys (CSUR)},
  vol.~51, no.~4, pp.~1--36, 2018.

\bibitem{li2023attack}
J.~Li, Z.~Wang, Y.~Shen, and L.~Xie, ``Attack detection for cyber-physical
  systems: A zonotopic approach,'' {\em IEEE Transactions on Automatic
  Control}, 2023.

\bibitem{hashemi2019co}
N.~Hashemi and J.~Ruths, ``Co-design for resilience and performance,'' {\em
  IEEE Transactions on Control of Network Systems}, pp.~1--12, 2022.

\bibitem{hespanha2019output}
J.~P. Hespanha and S.~D. Bopardikar, ``Output-feedback linear quadratic robust
  control under actuation and deception attacks,'' in {\em 2019 Am. Control
  Conference (ACC)}, pp.~489--496, IEEE, 2019.

\bibitem{fotiadis2022concurrent}
F.~Fotiadis and K.~G. Vamvoudakis, ``Concurrent receding horizon control and
  estimation against stealthy attacks,'' {\em IEEE Transactions on Automatic
  Control}, 2022.

\bibitem{milovsevic2020security}
J.~Milo{\v{s}}evi{\'c}, A.~Teixeira, T.~Tanaka, K.~H. Johansson, and
  H.~Sandberg, ``Security measure allocation for industrial control systems:
  Exploiting systematic search techniques and submodularity,'' {\em
  International Journal of Robust and Nonlinear Control}, vol.~30, no.~11,
  pp.~4278--4302, 2020.

\bibitem{tom2008recommended}
S.~Tom, D.~Christiansen, and D.~Berrett, ``Recommended practice for patch
  management of control systems,'' tech. rep., Idaho National Lab.(INL), Idaho
  Falls, ID (United States), 2008.

\bibitem{beattie2002timing}
S.~Beattie, S.~Arnold, C.~Cowan, P.~Wagle, C.~Wright, and A.~Shostack, ``Timing
  the application of security patches for optimal uptime.,'' in {\em LISA},
  vol.~2, pp.~233--242, 2002.

\bibitem{risley2003electronic}
A.~Risley, J.~Roberts, and P.~LaDow, ``Electronic security of real-time
  protection and {SCADA} communications,'' {\em Schweitzer Engineering
  Laboratories, SEL}, 2003.

\bibitem{gomez2018operator}
M.~Ekstedt, {\em Operator authentication and accountability for SCADA servers
  when requests are forwarded by a middle layer}.
\newblock PhD thesis, Aalto University, 2018.

\bibitem{hadley2007secure}
M.~Hadley and K.~Huston, ``Secure {SCADA} communication protocol performance
  test results,'' {\em Pacific Northwest National Laboratory (August 2007)},
  2007.

\bibitem{mo2009secure}
Y.~Mo and B.~Sinopoli, ``Secure control against replay attacks,'' in {\em 2009
  47th annual Allerton conference on communication, control, and computing
  (Allerton)}, pp.~911--918, IEEE, 2009.

\bibitem{ferrari2020switching}
R.~M. Ferrari and A.~M. Teixeira, ``A switching multiplicative watermarking
  scheme for detection of stealthy cyber-attacks,'' {\em IEEE Trans. on
  Automatic Control}, vol.~66, no.~6, pp.~2558--2573, 2020.

\bibitem{du2021secure}
D.~Du, C.~Zhang, X.~Li, M.~Fei, T.~Yang, and H.~Zhou, ``Secure control of
  networked control systems using dynamic watermarking,'' {\em IEEE Trans. on
  Cybernetics}, vol.~52, no.~12, pp.~13609--13622, 2021.

\bibitem{dan2010stealth}
G.~D{\'a}n and H.~Sandberg, ``Stealth attacks and protection schemes for state
  estimators in power systems,'' in {\em 2010 first IEEE {I}nternational
  {C}onference on {S}mart {G}rid {C}ommunications}, pp.~214--219, IEEE, 2010.

\bibitem{nguyen2022zero}
A.~T. Nguyen, S.~C. Anand, and A.~M. Teixeira, ``A zero-sum game framework for
  optimal sensor placement in uncertain networked control systems under
  cyber-attacks,'' in {\em 2022 IEEE 61st Conference on Decision and Control
  (CDC)}, pp.~6126--6133, IEEE, 2022.

\bibitem{rockafellar2000optimization}
R.~T. Rockafellar and S.~Uryasev, ``Optimization of conditional
  value-at-risk,'' {\em Journal of risk}, vol.~2, pp.~21--42, 2000.

\bibitem{krause2011randomized}
A.~Krause, A.~Roper, and D.~Golovin, ``Randomized sensing in adversarial
  environments,'' in {\em Twenty-Second International Joint Conference on
  Artificial Intelligence}, 2011.

\bibitem{rahmattalabi2018robust}
A.~Rahmattalabi, P.~Vayanos, and M.~Tambe, ``A robust optimization approach to
  designing near-optimal strategies for constant-sum monitoring games,'' in
  {\em Decision and Game Theory for Security: 9th International Conference,
  GameSec 2018, Seattle, WA, USA, October 29--31, 2018, Proceedings 9},
  pp.~603--622, Springer, 2018.

\bibitem{pirani2021game}
M.~Pirani, E.~Nekouei, H.~Sandberg, and K.~H. Johansson, ``A game-theoretic
  framework for security-aware sensor placement problem in networked control
  systems,'' {\em IEEE Transactions on Automatic Control}, vol.~67, no.~7,
  pp.~3699--3706, 2021.

\bibitem{pirani2021strategic}
M.~Pirani, J.~A. Taylor, and B.~Sinopoli, ``Strategic sensor placement on
  graphs,'' {\em Systems \& Control Letters}, vol.~148, p.~104855, 2021.

\bibitem{milosevic2023strategic}
J.~Milosevic, M.~Dahan, S.~Amin, and H.~Sandberg, ``Strategic monitoring of
  networked systems with heterogeneous security levels,'' {\em arXiv preprint
  arXiv:2304.04131}, 2023.

\bibitem{milovsevic2019network}
J.~Milo{\v{s}}evi{\'c}, M.~Dahan, S.~Amin, and H.~Sandberg, ``A network
  monitoring game with heterogeneous component criticality levels,'' in {\em
  2019 IEEE 58th Conference on Decision and Control (CDC)}, pp.~4379--4384,
  IEEE, 2019.

\bibitem{anand2020joint}
S.~C. Anand and A.~M. Teixeira, ``Joint controller and detector design against
  data injection attacks on actuators,'' {\em IFAC-PapersOnLine}, vol.~53,
  no.~2, pp.~7439--7445, 2020.

\bibitem{anand2022risk_acc}
S.~C. Anand and A.~M. Teixeira, ``Risk-averse controller design against data
  injection attacks on actuators for uncertain control systems,'' in {\em 2022
  Am. Control Conference (ACC)}, pp.~5037--5042, IEEE, 2022.

\bibitem{anand2022risk}
S.~C. Anand, A.~M. Teixeira, and A.~Ahl{\'e}n, ``Risk assessment and optimal
  allocation of security measures under stealthy false data injection
  attacks,'' in {\em 2022 IEEE Conference on Control Technology and
  Applications (CCTA)}, pp.~1347--1353, IEEE, 2022.

\bibitem{tan2009unified}
W.~Tan, ``Unified tuning of pid load frequency controller for power systems via
  {IMC},'' {\em IEEE Trans. on power systems}, vol.~25, no.~1, pp.~341--350,
  2009.

\bibitem{verhaegen2007filtering}
M.~Verhaegen and V.~Verdult, {\em Filtering and system identification: a least
  squares approach}.
\newblock Cambridge university press, 2007.

\bibitem{gally2018framework}
T.~Gally, M.~E. Pfetsch, and S.~Ulbrich, ``A framework for solving
  mixed-integer semidefinite programs,'' {\em Optimization Methods and
  Software}, vol.~33, no.~3, pp.~594--632, 2018.

\bibitem{anand2021risk}
S.~C. Anand, A.~M. Teixeira, and A.~Ahl{\'e}n, ``Risk assessment of stealthy
  attacks on uncertain control systems,'' {\em arXiv preprint
  arXiv:2106.07071}, 2021.

\bibitem{trentelman1991dissipation}
H.~L. Trentelman and J.~C. Willems, ``The dissipation inequality and the
  algebraic riccati equation,'' in {\em The Riccati Equation}, pp.~197--242,
  Springer, 1991.

\bibitem{petersen2012robust}
I.~R. Petersen, V.~A. Ugrinovskii, and A.~V. Savkin, {\em Robust Control Design
  Using $H_{\infty}$ Methods}.
\newblock Springer Science \& Business Media, 2012.

\bibitem{muller2018risk}
M.~I. M{\"u}ller, J.~Milo{\v{s}}evi{\'c}, H.~Sandberg, and C.~R. Rojas, ``A
  risk-theoretical approach to $\mathcal{H}_2$-optimal control under covert
  attacks,'' in {\em 2018 IEEE Conf on Decision and Contr (CDC)},
  pp.~4553--4558, IEEE.

\bibitem{hilhorst2014extended}
G.~Hilhorst, G.~Pipeleers, R.~C. Oliveira, P.~L. Peres, and J.~Swevers, ``On
  extended {LMI} conditions for ${H}_{2}/{H}_{\infty}$ control of {DT} linear
  systems,'' {\em IFAC Proc. Volumes}, vol.~47, no.~3, pp.~9307--9312, 2014.

\bibitem{bopardikar2016h}
S.~D. Bopardikar, A.~Speranzon, and J.~P. Hespanha, ``An ${H}_{\infty}$
  approach to stealth-resilient control design,'' in {\em 2016 Resilience Week
  (RWS)}, pp.~56--61, IEEE, 2016.

\bibitem{you2015convex}
S.~You and N.~Matni, ``A convex approach to sparse ${H}_{\infty}$ analysis \&
  synthesis,'' in {\em 2015 54th IEEE Conference on Decision and Control
  (CDC)}, pp.~6635--6642, IEEE, 2015.

\bibitem{yuan2016binary}
G.~Yuan and B.~Ghanem, ``Binary optimization via mathematical programming with
  equilibrium constraints,'' {\em arXiv preprint arXiv:1608.04425}, 2016.

\bibitem{fischetti2010heuristics}
M.~Fischetti and A.~Lodi, ``Heuristics in mixed integer programming,'' {\em
  Wiley encyclopedia of operations research and management science}, 2010.

\bibitem{muller2022risk}
M.~I. M{\"u}ller and C.~R. Rojas, ``Risk-theoretic optimal design of
  output-feedback controllers via iterative convex relaxations,'' {\em
  Automatica}, vol.~136, p.~110042, 2022.

\bibitem{wolsey1982analysis}
L.~A. Wolsey, ``An analysis of the greedy algorithm for the submodular set
  covering problem,'' {\em Combinatorica}, vol.~2, no.~4, pp.~385--393, 1982.

\bibitem{goodwin2014adaptive}
G.~C. Goodwin and K.~S. Sin, {\em Adaptive filtering prediction and control}.
\newblock Courier Corporation, 2014.

\bibitem{caverly2019lmi}
R.~J. Caverly and J.~R. Forbes, ``{LMI} properties and applications in systems,
  stability, and control theory,'' {\em arXiv preprint arXiv:1903.08599}, 2019.

\end{thebibliography}
\end{document}